\documentclass[sigconf,nonacm,balance=false]{acmart}
\usepackage{popets}
\AtBeginDocument{%
  }

\usepackage{subcaption}

\usepackage{xcolor}
\usepackage{multirow}
\usepackage[flushleft]{threeparttable}
\usepackage{amsmath}
\usepackage{pifont}
\newcommand{\cmark}{\ding{51}}
\newcommand{\xmark}{\ding{55}}

\usepackage{algorithmic}
\usepackage[ruled,vlined,linesnumbered]{algorithm2e}

\usepackage[T1]{fontenc}
\usepackage{graphicx}
\usepackage{booktabs}

\begin{document}

\title{Examining the Rat in the Tunnel: Interpretable Multi-Label Classification of Tor-based Malware}

\author{Ishan Karunanayake}
\affiliation{%
  \institution{University of New South Wales}
  \city{Sydney}
  \country{Australia}
  }
\affiliation{%
  \institution{Cyber Security Cooperative Research Centre (CSCRC)}
  \city{Joondalup}
  \country{Australia}
  }
\email{ishan.karunanayake@unsw.edu.au}

\author{Mashael AlSabah}
\affiliation{%
  \institution{Qatar Computing Research Institute}
  \city{Ar-Rayyan}
  \country{Qatar}
  }

\author{Nadeem Ahmed}
\affiliation{%
  \institution{University of New South Wales}
  \city{Sydney}
  \country{Australia}
  }
\affiliation{%
  \institution{Cyber Security Cooperative Research Centre (CSCRC)}
  \city{Joondalup}
  \country{Australia}
  }

\author{Sanjay Jha}
\affiliation{%
  \institution{University of New South Wales}
  \city{Sydney}
  \country{Australia}
  }
\affiliation{%
  \institution{Cyber Security Cooperative Research Centre (CSCRC)}
  \city{Joondalup}
  \country{Australia}
  }

\begin{abstract}
Despite being the most popular privacy-enhancing network, Tor is increasingly adopted by cybercriminals to obfuscate malicious traffic, hindering the identification of malware-related communications between compromised devices and Command and Control (C\&C) servers. This malicious traffic can induce congestion and reduce Tor's performance, while encouraging network administrators to block Tor traffic. Recent research, however, demonstrates the potential for accurately classifying captured Tor traffic as malicious or benign. While existing efforts have addressed malware class identification, their performance remains limited, with micro-average precision and recall values around 70\%. Accurately classifying specific malware classes is crucial for effective attack prevention and mitigation. Furthermore, understanding the unique patterns and attack vectors employed by different malware classes helps the development of robust and adaptable defence mechanisms. 

We utilise a multi-label classification technique based on Message-Passing Neural Networks, demonstrating its superiority over previous approaches such as Binary Relevance, Classifier Chains, and Label Powerset, by achieving micro-average precision (MAP) and recall (MAR) exceeding 90\%. Compared to previous work, we significantly improve performance by 19.98\%, 10.15\%, and 59.21\% in MAP, MAR, and Hamming Loss, respectively. Next, we employ Explainable Artificial Intelligence (XAI) techniques to interpret the decision-making process within these models. Finally, we assess the robustness of all techniques by crafting adversarial perturbations capable of manipulating classifier predictions and generating false positives and negatives.
\end{abstract}

\maketitle

\section{INTRODUCTION}

With growing worries about online privacy and security, people are increasingly turning to tools that hide their identity, like anonymity networks. Unfortunately, the Tor network, with its readily available anonymity features, has attracted cybercriminals seeking to conduct malicious activities. Further, the growing use of Tor by botnets and self-propagating malware, as evidenced by Mirai \cite{MIRAI_TOR_2019}, Brickerbot \cite{BRICKERBOT_2017}, and Wannacry \cite{WANNACRY_TOR}, poses a considerable threat due to the difficulty in tracking, locating, and dismantling such botnets. The negative perception of Tor due to this association with malware can reduce user confidence in its ability to provide privacy and anonymity. Additionally, the presence of malware and botnet traffic within the Tor network can lead to performance degradation due to congestion \cite{HOPPER2013}. These factors can motivate security professionals and network administrators to implement controls and restrictions on Tor traffic. Although malware and botnet detection has been a long-standing research field with a substantial body of work \cite{WANG2011, MASUD2008, GU2008_BOTSNIFFER, GU2009, ZENG2010, ZHAO2013, STEVANOVIC2014, MCDERMOTT2018}, research dedicated specifically to identifying Tor-based malware remains comparatively limited \cite{LING2015, FAJANA2018, DODIA2022}. The growing use of this method by criminals makes it even more urgent to investigate it further.

Ling et al. \cite{LING2015} deployed an Intrusion Detection System (IDS) adjacent to a Tor exit router and observed that 10\% of Tor traffic triggered IDS alerts, suggesting the presence of malware communications masked within Tor traffic. Subsequent efforts by Fajana et al. \cite{FAJANA2018} and Dodia et al. \cite{DODIA2022} employed traffic analysis techniques to successfully detect malware-embedded Tor traffic. Notably, Dodia et al. implemented the most rigorous data collection and evaluation process, demonstrating the ability to classify a given Tor traffic trace as benign or malicious with a maximum precision of 93.33\% and a maximum recall of 81.6\%. Furthermore, they attempted to infer the malware class (e.g., virus, downloader, miner, grayware) of the traffic trace. However, their best result using the Label Powerset technique yielded a micro-average precision of 66.81\%, a micro-average recall of 72.37\%, and a Hamming loss of 0.1.

\textbf{Motivation}: The accurate identification of malware classes holds critical importance for several reasons. The objectives, targets, attack vectors, and severity levels can vary significantly between different malware classes. Consequently, accurate classification facilitates the development of targeted countermeasures, incident response plans, and enhanced defensive mechanisms. Consider two scenarios: a system under attack by known and blocked malware versus one encountering a novel threat. Employing the same response and resource allocation in both situations could incur substantial costs for the affected organisation. Furthermore, identifying the malware class after an attack allows security teams to proactively develop future preventative measures. Despite its significance, malware class classification remains extremely challenging. Encrypted malware traffic over Tor further exacerbates detection and classification. Existing tools often lack sufficient information to effectively classify all malware variants, as evidenced by the extensive number of unknown malware binaries in the dataset employed by \cite{DODIA2022} (Figure \ref{fig:label_cooccur}). The scarcity of readily available datasets for Tor-based malware further makes it a challenging problem. Additionally, multi-label classification, employed in \cite{DODIA2022} presents inherent difficulties due to potential label noise and insufficient data samples. Our work directly addresses the problem of malware class identification for Tor-based malware and demonstrates that the utilisation of advanced neural network-based techniques can significantly improve classification performance.

\textbf{Explainability}: Existing traffic analysis research, such as Dodia et al. \cite{DODIA2022} and Fajana et al. \cite{FAJANA2018} followed a predominantly black-box approach to machine learning models. These models often exhibit acceptable performance without offering insights into their decision-making processes. Furthermore, rigorous evaluations, particularly on their robustness to adversarial manipulation, seem to be lacking. Although several of these models present promising results, their susceptibility to evasion through data manipulation remains a significant concern. To address this critical gap, we delve into a comprehensive analysis of feature importance and its influence on model predictions. Leveraging Explainable Artificial Intelligence (XAI) techniques within our experiments, we gain an in-depth understanding of the model's internal workings. Based on that, we design and implement targeted yet simple attacks capable of successfully evading many high-performing models.

We discuss the contributions of this paper by focusing on the following three research questions.

\textbf{1. RQ1: Is it possible to improve the classification performance for identifying malware classes in Tor traffic?}

Dodia et al. collected malware traces from the Tor network and labelled their dataset into nine distinct malware classes. They employed three widely used multi-label classification techniques\footnote{As a result of their labelling technique, a single trace may contain multiple (1-4) class labels. Consequently, this task requires the use of multi-label classification techniques. It is crucial to note that multi-label classification is distinct from multi-class classification, in which each instance is assigned only one label.} (Binary Relevance (BR) \cite{Zhang2018}, Classifier Chains (CC) \cite{Read2011}, and Label Powerset (LP) \cite{TSOUMAKAS2007}) to classify different traffic traces in their dataset. Of these techniques, LP achieved the best results. In this paper, we explore state-of-the-art machine learning techniques to evaluate whether we can improve the classification performance.

\textbf{Contribution}: We employ a Message Passing Neural Network (MPNN)-based technique that achieves higher accuracy for malware class identification. We use five datasets (D5, D10, D20, D30, and D40) containing varying numbers of samples for evaluation (generated by different numbers of malware binaries as explained in Section \ref{sec:dataset}). Additionally, we compare this technique with previously reported techniques in \cite{DODIA2022}. The MPNN-based technique significantly outperformed the other classifiers, achieving a 95.44\% Micro Average Precision (MAP), 93.98\% Micro Average Recall (MAR), 0.018 Hamming Loss (HL), and 88.46\% accuracy on the D20 dataset. This is an improvement of 27\% in MAP, 22.08\% in MAR, 78.05\% in HL, and 65.69\% in accuracy compared to the LP technique, which showed the second-best performance (Table \ref{tab:rq1_results}).

\textbf{2. RQ2: Are we able to see what factors led the models to identify a specific malware class?}

Machine learning models are often treated as black boxes, with their results being difficult to interpret. To be widely adopted, these techniques need to be able to explain their decision-making process.
Recognising the importance of understanding how machine learning models reach their decisions, the field of Explainable Artificial Intelligence (XAI) emerged \cite{Holzinger2022}. In this work, we employ XAI techniques, specifically, Shapley Additive Explanations (SHAP), to interpret classifier predictions. SHAP allows us to not only identify important features but also to understand how each feature affects any given prediction. We use these insights to understand the similarities and differences between malware classes.

\textbf{Contribution}: We identify how various features affect the predictions for different classes. We provide visualisation via SHAP plots to interpret model predictions, including summary plots for global interpretation (understanding model outcomes as a whole), as well as force plots and decision plots for local interpretation (understanding the predictions of individual samples). We observed that the performance of BR, CC, and LP techniques is particularly affected by the lack of data for less common malware types like Keylogger, Worm, Spyware, and Backdoor.

\textbf{3. RQ3: Is it possible to create variations of malware traffic (adversarial perturbations) that bypass our trained model and evade detection?}

A well known weakness of machine learning models is their susceptibility to adversarial attacks. This study explores if we can exploit the unique characteristics of different malware types (what makes them similar and different) to create adversarial perturbations in the traffic patterns. Our goal is to see if these altered patterns can fool our detection models.
As an example, we alter Ransomware traffic to mimic Downloader traffic. Using the same methods to understand the model's decisions (XAI techniques), we can see how surprisingly simple changes, like adjusting the duration of a Tor connection and the number of data packets sent, can trick the system into mistakenly identifying malware (false positives) or missing it entirely (false negatives).

\textbf{Contribution}: Our research  showed that we could use explanations from SHAP plots to create "adversarial examples" that trick the malware classifier. We did this by finding features that make the classifier think it's Downloader traffic and not Ransomware, then modifying real Ransomware samples. This fooled the simpler classifiers, causing them to wrongly identify Ransomware as Downloader or even the less harmful Grayware. However, the more advanced model based on MPNNs shows better robustness and an ability to maintain relatively high Ransomware predictions and low Downloader and Grayware predictions (Table \ref{tab:evasion_attack_results}).

The rest of this paper is organised as follows. Section \ref{sec:related_work} establishes the necessary context by reviewing relevant background and existing work in the field. Section \ref{sec:dataset} then introduces the specific dataset employed in our investigation. The subsequent sections, \ref{sec:rq1}, \ref{sec:rq2}, and \ref{sec:rq3}, each address one of the three research questions outlined previously, presenting a detailed account of the experimental design, results obtained, and subsequent analysis. Finally, Section \ref{sec:conclusion} concludes with insights on potential future work.


\section{BACKGROUND AND RELATED WORK}
\label{sec:related_work}
In this section, we discuss the prior research related to this work.

\begin{table*}[h]
  \centering
  \caption{Comparing Related Work on Tor-based Malware}
  \tabcolsep=0.09cm
  \scalebox{0.9}{
  \begin{tabular}{l|p{3.5cm}|p{3.5cm}|p{5cm}|c|c}
    \toprule
    Publication & Main Focus & Approach & Performance & Interpretability & Evaluate Robustness\\
    \midrule
    Ling et al. \cite{LING2015} & Detect malicious Tor traffic & Configuring an IDS behind an exit router & 10\% of the exit router traffic is malicious & N/A & N/A \\
    Fajana et al. \cite{FAJANA2018} & Detect malicious Tor traffic & Using ML on circuit-level data from entry nodes & 99.6\% accuracy with Random Forest & \xmark & \xmark\\
    Dodia et al. \cite{DODIA2022} & (1) Detect malicious Tor traffic and (2) Malware class classification & Encrypted traffic analysis & (1) LightGBM: 93.33\% precision \& 81.6\% recall (2) Label Powerset: 66.81\% MAP, 72.37\% MAR, 0.1 HL  & \xmark & \xmark\\
    This work & Malware class classification & Encrypted traffic analysis & MAP>92\%, MAR>90\%, HL=<0.031 &  \cmark & \cmark\\
    \bottomrule
  \end{tabular}}
  \label{tab:prior_work_comparison}
\end{table*}

\subsection{The Tor network and Hidden Services}
The Tor network leverages multiple relays to ensure user anonymity. When users connect to web services through Tor, their traffic is encrypted and routed through a series of intermediary relays, including entry, middle, and exit relays. Adversaries who intercept traffic between any two points of a Tor circuit cannot associate the endpoints. While this setup anonymises users, it does not anonymise web services, as users can easily discover their IP addresses. To address this, Tor introduced hidden services, also known as onion services. These web services are only accessible via the Tor network, and their actual IP addresses are concealed from users. To access a hidden service, users need a Tor client, which retrieves the service descriptor from a Hidden Service Directory (HSDir) and establishes a circuit to the service through a rendezvous point. This approach grants anonymity to both users and hidden services \cite{DINGLEDINE2004tornetwork}.

\subsection{Passive analysis of Tor traffic}
Tor's robust anonymity has positioned it as a target for numerous traffic analysis attacks. While Tor's encryption thwarts direct packet content inspection, research indicates that metadata (e.g., packet size, direction, and timestamp) can be leveraged to extract fingerprints from traffic traces—defined as sets of consecutive packets with identical source/destination IP addresses and ports, and protocol—and subsequently train supervised machine learning classifiers for identifying similar traces. This passive traffic analysis technique has primarily been employed for two purposes against Tor, as outlined below.

\subsubsection{Classify Tor traffic from other types of traffic}
While Tor's anonymity facilitates censorship circumvention, attempts to restrict its use have emerged. Blocking all publicly listed Tor nodes is one such approach. However, it is ineffective due to the network's bridge relays, which offer unlisted access points. Passive traffic analysis offers a potential alternative. Lashkari et al. \cite{LASHKARI2017} and Montieri et al. \cite{Montieri2020} demonstrated the feasibility of classifying Tor traffic via manually extracted statistical features (e.g., packet inter-arrival times) and machine learning models (e.g., Random Forests, Decision Trees). These findings indicate the possibility of distinguishing Tor traffic not only from regular Internet traffic but also from other anonymity networks like I2P and JonDonym.

\subsubsection{Website Fingerprinting}
Website fingerprinting aims to identify user activity within anonymity networks by analysing traffic metadata. Attackers capture client traffic towards the entry guard and generate fingerprints based on this data. These fingerprints are then used to train machine-learning models for website identification. However, capturing fingerprints for all websites is impractical. Therefore, a two-phase approach is employed: first, a model distinguishes the user's traffic as belonging to a pre-defined set of monitored websites; if so, a second model further identifies the specific website within that set. Extensive research has explored website fingerprinting in the context of Tor, demonstrating its potential and limitations \cite{SIRINAM2018, RIMMER2018, PANCHENKO2016, HAYES2016, WANG2014, BHAT2019, RAHMAN2020}.

\subsection{Tor-based Malware}

\begin{figure}[h]
\centering
\includegraphics[width=\linewidth]{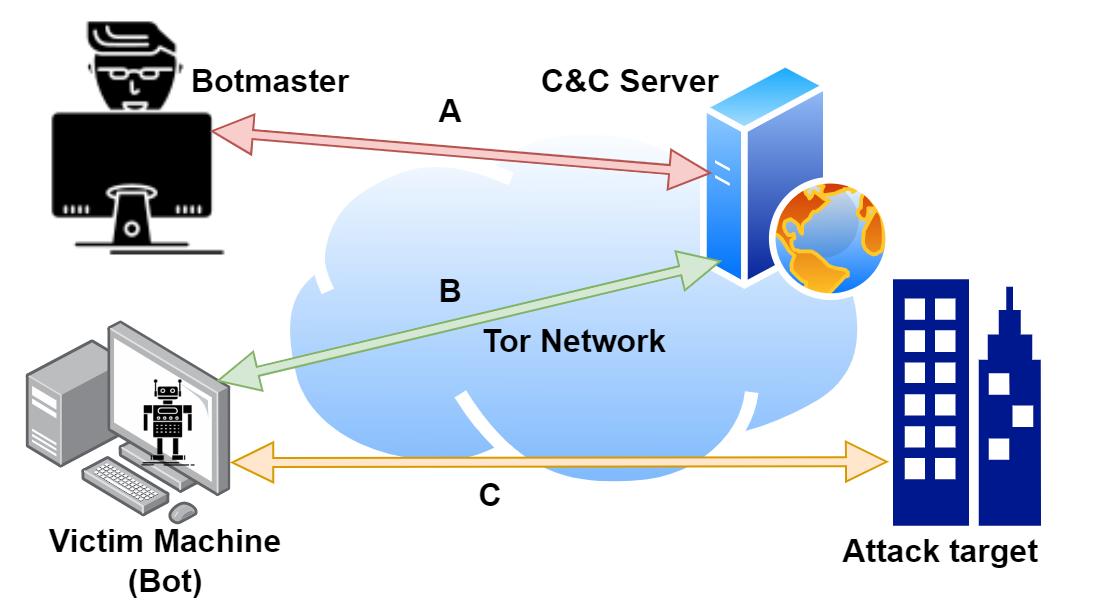}
\caption{How Malware can use Tor. (A) The botmaster can access the C\&C server (hidden service or not) via Tor. (B) C\&C and its victims (bots) communicate via Tor. (C) Infected machines (bots) can propagate the malware or attack other networks via Tor.}
\label{fig:tor_malware}
\end{figure}

There are a number of ways that a botnet or malware can use the Tor network to its advantage. Figure \ref{fig:tor_malware} illustrates some general use cases of Tor for a botnet. The Tor network can mainly be used for communication between all entities of a botnet. Additionally, the C\&C server can be hosted as a hidden service to provide an additional layer of defence for the botnet. All of these components make it more difficult for cybersecurity professionals and law enforcement teams to prevent malware propagation, shut down C\&C servers, and apprehend cybercriminals.

There has been limited research on identifying Tor-based malware. A major reason for this is the lack of high-quality data that can be used in experiments. However, Ling et al. \cite{LING2015} introduced TorWard, a system that can detect, classify, trace back, and block malicious Tor traffic. TorWard is the first system to attempt to categorise malicious traffic over Tor. The authors found that approximately 10\% of Tor traffic contains malicious content that can trigger IDS alerts. TorWard avoids administrative and legal complaints arising from this malicious content by redirecting the traffic back into the Tor network. However, TorWard has two main drawbacks. First, its performance is heavily dependent on the performance of the IDS. Second, it is ineffective if the C\&C server is a hidden service. Fajana et al. \cite{FAJANA2018} proposed a traffic classification-based technique for detecting Tor botnets called \textit{TorBot Stalker}. TorBot Stalker extracts circuit-level data from entry nodes and uses it to train machine learning classifiers that can detect malware. The authors' experimental results were promising, with an accuracy of 99\% and very few false positives. However, TorBot Stalker has two main drawbacks. First, it must be implemented by a substantial number of entry nodes to be effective in the live network. Second, it cannot identify periodic patterns in short-lived connections.

Dodia et al. \cite{DODIA2022} collected an extensive dataset of Tor-based malware and made it publicly available (more details about the dataset will be presented in Section \ref{sec:dataset}). The authors first attempted to identify whether a given Tor traffic trace contained malware or not using a binary classification model. Second, they attempted to identify the malware class of the malware traces using multi-label classification techniques. Multi-label classification was employed because some malware traces fell under two or more classes (as a result of their labelling technique). In their experiments, Dodia et al. used two types of features: connection-level features and host-level features. Connection-level features were the same features introduced by Hayes et al. \cite{HAYES2016} for a WF attack, whereas host-level features were newly introduced. For the binary classification task, the authors used the AutoGluon \cite{AUTOGLUON} library and found that the LightGBM model performed best, with a maximum precision of 93.33\% and a maximum recall of 81.6\%. For the multi-label classification problem, the authors used the \texttt{scikit-learn} and \texttt{scikit-multilearn} libraries for experiments. Out of the three multi-label classification techniques (Binary Relevance, Classifier Chains, and Label Powerset) that the authors employed with a Random Forest base classifier, ``Label Powerset'' performed best with a 66.81\% MAP and a 72.37\% MAR (these metrics will be explained later in Section \ref{sec:rq1}). Table \ref{tab:prior_work_comparison} compares the existing work that focus on Tor-based malware and highlights the unique contributions of our work.


\section{DATASET}
\label{sec:dataset}
As previously mentioned, this work uses the publicly available dataset from \cite{DODIA2022}. While we refer interested readers to the original paper for detailed information on the data collection process, this section provides some statistics relevant to our work that were not discussed in the previous work.

\begin{table}[h]
  \centering
  \caption{Malware Traffic Datasets used in \cite{DODIA2022}}
  \tabcolsep=0.09cm
  \scalebox{0.9}{
  \begin{tabular}{c|c|c|c}
    \toprule
    Dataset & No. of binaries & PCAPs per binary & Total classifier instances\\
    \midrule
    D5 & 157 & 5 & 2,027 \\
    D10 & 130 & 10 & 3,657\\
    D20 & 107 & 20 & 6,135\\
    D30 & 62 & 30 & 5,342 \\
    D40 & 17 & 40 & 1,940 \\
    \bottomrule
  \end{tabular}}
  \label{tab:different_datasets}
\end{table}

\begin{table}[h]
  \centering
  \caption{Malware Class Distribution}
  \scalebox{0.9}{
  \begin{tabular}{l|r|r|r|r|r}
    \toprule
    Malware Class & D5 & D10 & D20 & D30 & D40\\
    \midrule
    Grayware & 1,263 & 2,294 & 3,686 & 2,940 & 1,138  \\
    Downloader & 1,186 & 2,161 & 3,372 & 2,543 & 347  \\
    Ransomware & 365 & 729 & 1,455 & 2,003 & 900  \\
    Miner & 384 & 625 & 550 & 174 & 0  \\
    Virus & 59 & 118 & 234 & 174 & 232  \\
    Spyware & 45 & 87 & 116 & 175 & 0  \\
    Backdoor & 59 & 117 & 236 & 90 & 120  \\
    Keylogger & 15 & 28 & 58 & 87 & 0  \\
    Worm & 75 & 48 & 91 & 0 & 0  \\
    Unknown & 179 & 319 & 591 & 615 & 235  \\
    \bottomrule
  \end{tabular}}
  \label{tab:class_distribution}
\end{table}

\begin{figure}[h]
\centering
\includegraphics[width=\linewidth]{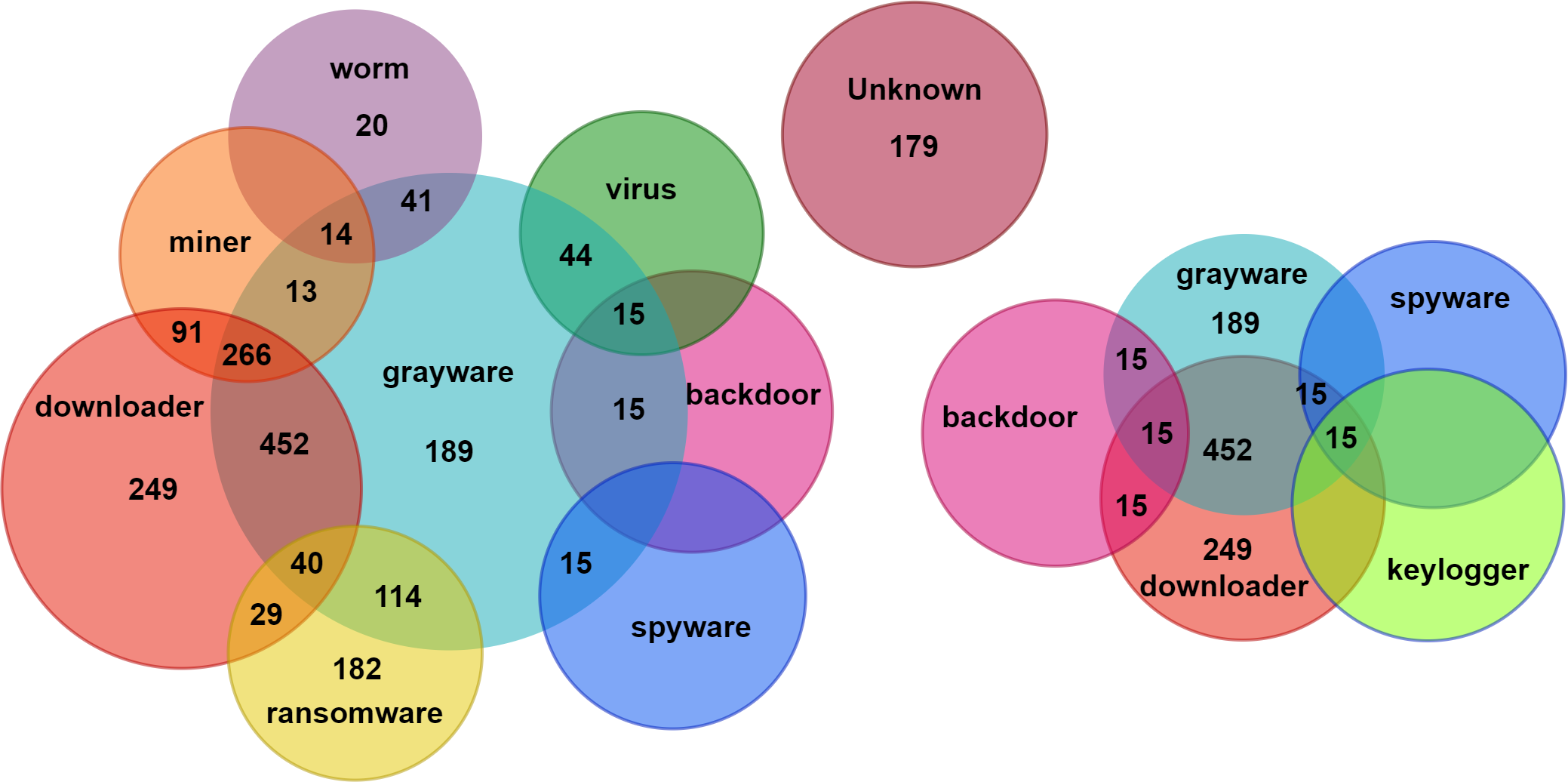}
\caption{Label Co-occurrence Diagram for the D5 dataset. The diagram on the left provides a more wholesome view of the label co-occurrences, while the diagram on the right shows additional co-occurrences that cannot be depicted in the left diagram. The unknown circle is isolated as it represents unidentified malware classes.}
\label{fig:label_cooccur}
\end{figure}

To gain better insights into the machine learning model's predictions, we need to have a good understanding of the datasets. The dataset contains 5,984 malware PCAPs generated by 362 malware binaries. Those are organised in five experimental datasets as described in Table \ref{tab:different_datasets}. Table \ref{tab:class_distribution} shows the number of classifier instances of each class in each dataset, where a \textit{classifier instance} refers to a single data point within the dataset. Figure \ref{fig:label_cooccur} provides an overview of how these labels co-occur in the \textit{D5} dataset. Other datasets contain subsets of these co-occurrences. For example, \textit{D10} dataset does not contain instances of \textit{Grayware} and \textit{Worm} co-occurring (without \textit{miner}), while \textit{D20} does not contain (\textit{Grayware + Worm}) and (\textit{Grayware + Spyware}) traces. Overall, there are 22 different combinations of co-occurrences in \textit{D5} and \textit{D10}, \textit{D20}, \textit{D30}, and \textit{D40} contains 21, 20, 14, and 8 out of those 22, respectively.


\section{Identifying Malware Classes}
\label{sec:rq1}
In this section, we address \textbf{\textit{RQ1: Is it possible to improve the classification performance for identifying malware classes in Tor traffic?}} First, we review the technique we have used and then conduct a comprehensive evaluation of all techniques, including the state-of-the-art.

Dodia et al. presented the first set of experiments conducted on their dataset to identify different malware classes. In their experiments, they used three multi-label classification techniques - Binary Relevance \cite{Zhang2018}, Classifier Chains \cite{Read2011}, and Label Powerset \cite{TSOUMAKAS2007} - with a Random Forest base model. We use these techniques for comparison in our work.

\subsection{Label Message Passing}
After reviewing existing multi-label classification techniques, we identified Label Message Passing (LaMP) \cite{LANCHANTIN2020}, a technique based on Message Passing Neural Networks (MPNN). Our preliminary experiments showed that LaMP can achieve superior results for our problem.

\subsubsection{Message Passing Neural Network}

MPNN is a type of neural network designed to process and exchange information between nodes in a graph. This network is especially useful for tasks involving structured data, where there are relationships between different entities. The main idea is to pass messages between nodes and update their states iteratively.

Let us consider a graph $G$ with a set of nodes $V$. At iteration $t$, node $v_i$ ($v_i \in V$) exchanges information (messages) with its neighbouring nodes and updates its representations. The message from node $v_j$ ($v_j \in V$) to node $v_i$ at step $t$ denoted by $m_{ij}^{(t)}$ can be calculated as:
\begin{equation}
    m_{ij}^{(t)} = M\left(h_i^{(t-1)}, h_j^{(t-1)}\right),
\end{equation}
where $M$ is a message function that computes messages based on the previous states of the connected nodes. Then, all the messages received by $v_i$ from its neighbours are aggregated. The aggregated message at node $v_i$ is denoted as $M_i^{(t)}$ and can be computed as:
\begin{equation}
    M_i^{(t)} = \sum_{j \in \text{neighbors of } v_i} m_{ij}^{(t)}
\end{equation}
Finally, the node $v_i$ updates its state $h_i^{(t)}$ using the aggregated message and its previous state:
\begin{equation}
    h_i^{(t)} = U\left(h_i^{(t-1)}, M_i^{(t)}\right),
\end{equation}
where $U$ is an update function that combines the previous state $h_i^{(t-1)}$ and the aggregated message $M_i^{(t)}$ to produce the updated state $h_i^{(t)}$ for node. These steps (message passing, aggregation, and state updating) are repeated for a fixed number of iterations or until convergence. After the desired number of iterations, we can use the final node representations $h_i^{(T)}$ (where $T$ is the number of iterations) for tasks like classification and regression.

\begin{figure}[h]
\centering
\includegraphics[width=\linewidth]{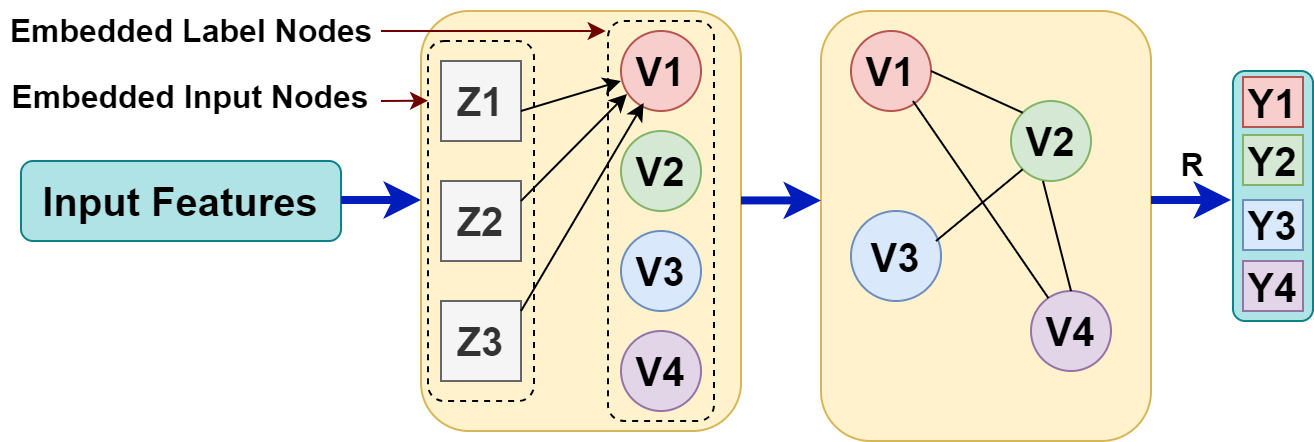}
\caption{Concept behind LaMP. A given set of input features is embedded into input nodes $\{Z1,Z2,Z3\}$ and the labels are embedded into label nodes $\{V1,V2,V3,V4\}$. The embedded label nodes are used to create a label interaction graph. First, messages are passed from input nodes to label nodes and update the label nodes. Then, messages are passed between labels before updating them again. After $T$ iterations, a readout function $R$ performs a node-level classification and makes binary predictions $\{Y1,Y2,Y3,Y4\}$ for each label.}
\label{fig:lamp_network}
\end{figure}

\subsubsection{Label Message Passing (LaMP)}
LaMP uses MPNNs on a Label Interaction Graph, as depicted in Figure \ref{fig:lamp_network}. \cite{LANCHANTIN2020} uses two main structures to create the Label Interaction Graph: (1) Prior: This structure uses a known or predefined label structure or creates edges between co-occurring labels. We use this structure in our experiments. (2) Fully connected: This structure connects all label nodes to each other. The overall model consists of two MPNNs, one for feature-to-label message passing and the other for label-to-label message passing. After a certain number of iterations, we can obtain a binary prediction for each label using a readout function.
We can compare these predictions to the actual predictions to determine the model's performance on that specific task. LaMP offers a distinct advantage over previous techniques – its ability to capture intricate label dependencies. Additionally, it allows us to tailor the graph structure based on the relationships between labels. However, a potential drawback of LaMP is its computational cost, which can become significant when dealing with a large number of labels.

\begin{table*}[h]
  \centering
  \caption{Evaluation of different multi-label classification techniques for malware class identification. MAP: Micro-Average Precision, MAR: Micro-Average Recall, HL: Hamming Loss, AC: Accuracy}
  \tabcolsep=0.09cm
  \scalebox{0.9}{
  \begin{tabular}{c|c|c|c|c|c|c|c|c|c|c|c|c|c|c|c|c}
    \toprule
    \multirow{2}{*}{Dataset} & \multicolumn{4}{c|}{Binary Relevance} & \multicolumn{4}{c|}{Classifier Chains} & \multicolumn{4}{c|}{Label Powerset} & \multicolumn{4}{c}{Label Message Passing}\\
    \cline{2-17}
    & MAP(\%) & MAR(\%) & HL & AC(\%) & MAP(\%) & MAR(\%) & HL & AC(\%) & MAP(\%) & MAR(\%) & HL & AC(\%) & MAP(\%) & MAR(\%) & HL & AC(\%) \\
    \midrule
    D5 & 75.79 & 75.85 & 0.086 & 44.51 & 75.77 & 76.67 & 0.085 & 45.96 & 76.69 & 82.33 & 0.076 & 56.98 & \textbf{92.01} & \textbf{90.69} & \textbf{0.031} & \textbf{78.36}\\
    D10 & 75.65 & 77.05 & 0.086 & 43.20 & 75.61 & 77.50 & 0.085 & 44.50 & 77.58 & 81.72 & 0.075 & 54.72 & \textbf{94.98} & \textbf{92.80} & \textbf{0.022} & \textbf{84.74}\\
    D20 & 73.81 & 73.95 & 0.088 & 42.80 & 74.00 & 74.29 & 0.087 & 44.46 & 75.15 & 76.98 & 0.082 & 53.39 & \textbf{95.44} & \textbf{93.98} & \textbf{0.018} & \textbf{88.46}\\
    D30 & 77.87 & 69.23 & 0.093 & 44.06 & 77.78 & 69.00 & 0.093 & 47.41 & 77.46 & 73.04 & 0.089 & 55.71 & \textbf{95.31} & \textbf{93.50} & \textbf{0.023} & \textbf{87.99}\\
    D40 & 77.53 & 69.41 & 0.130 & 47.13 & 77.27 & 72.92 & 0.124 & 54.09 & 80.07 & 77.67 & 0.106 & 69.09 & \textbf{95.56} & \textbf{92.33} & \textbf{0.031} & \textbf{86.69}\\
    \bottomrule
  \end{tabular}}
  \label{tab:rq1_results}
\end{table*}

\subsection{Evaluation}
This section details the experiments conducted to address our first research question. Initially, we sought to replicate the findings presented in \cite{DODIA2022} by employing their publicly available dataset and code. It is important to note that \cite{DODIA2022} only reported results for dataset D5. To further solidify our findings for RQ1, we conducted the experiments for all five datasets. Next, we assessed the performance of LaMP on these same datasets. In replicating the experiment, we maintained the same 70-30 train-test split utilised by \cite{DODIA2022}. Additionally, we mirrored their evaluation approach by employing three metrics – micro-average precision (MAP), micro-average recall (MAR), and Hamming loss (HL) – alongside accuracy (AC). we have provided more information about these metrics in the Appendix \ref{sec:appendix_A}.

While accuracy serves as a valuable metric for binary classification tasks, it falls short in multi-label settings. This limitation arises because when calculating accuracy, true positives, true negatives, false positives, and false negatives are calculated at the individual instance level. In multi-label problems, a single instance can encompass both correct and incorrect predictions. Consequently, a partially correct prediction holds more value than a completely incorrect one. Since accuracy fails to capture these nuances of partial correctness, metrics like HL become essential. However, a high accuracy score still indicates a model's ability to predict all labels for an instance accurately, representing a more robust measure of performance. Therefore, we have incorporated accuracy into our evaluation to showcase the superior performance of the MPNN-based model compared to previous techniques. Table \ref{tab:hyperparameters} shows the main hyperparameters used in the LaMP model.

\begin{table}[h]
  \centering
  \caption{Hyperparameters of the final model}
  \tabcolsep=0.09cm
  \scalebox{0.8}{
  \begin{tabular}{c|c|c}
    \toprule
    Parameter & Range & Final\\
    \midrule
    epochs & (10,20,30,40,50,100,200,300,500) & 100\\
    batch size & (16,32,64,128,256) & 64 \\
    optimizer & (adam, sgd) & adam \\
    learning rate & (0.001,0.0001,0.00001,0.0001-0.0005) & 0.0002 \\
    label mask & (none, prior) & prior\\
    dropout & (0.1,0.2,0.3,0.4,0.5) & 0.1 \\
    loss & (cross entrophy, adversarial, ranking) & cross entropy \\
    encoder & (mlp, graph, rnn) & graph \\
    decoder & (mlp, graph, rnn) & graph \\
    dimension of model & (64,128,256,512,1024)  & 512 \\
    hidden layer dimension & (64,128,256,512,1024,2048) & 1024\\
    \bottomrule
  \end{tabular}}
  \label{tab:hyperparameters}
\end{table}

\subsection{Results and Analysis}

Table \ref{tab:rq1_results} presents the results obtained in our experiments. We will discuss the interesting observations and insights from the table in the following section..

\textbf{Label Message Passing shows superior performance.} Table \ref{tab:rq1_results} reveals that the LaMP technique outperforms all three techniques previously used in \cite{DODIA2022} by a significant margin across all metrics. For all datasets, LaMP achieves MAP and MAR values greater than 90\%, with a maximum HL of 0.031. Additionally, the accuracy is greater than 80\% for four datasets, indicating that LaMP can correctly predict all labels for a given traffic trace more than 80\% of the time. The significance of these results can be further understood by comparing LaMP's performance to that of Label Powerset, the second-best technique. For example, on the D5 dataset, LaMP shows a 19.98\% improvement in MAP, a 10.15\% improvement in MAR, a 59.21\% improvement in HL, and a 37.52\% improvement in accuracy over Label Powerset. These gaps widen further for datasets such as D10 and D20. 

\textbf{Consistent results across different datasets.}
Another key observation from Table \ref{tab:rq1_results} is that while there are some improvements in the results when using different datasets, these changes are generally not very significant. To explore this statement in more detail, let us examine the results we obtained for LaMP in Table \ref{tab:rq1_results}. We can observe that LaMP achieves the best performance on the D20 dataset. According to Table \ref{tab:different_datasets}, the D20 dataset consists of 107 binary classes with 20 PCAPs per class. It achieves a MAP of 95.44\%, MAR of 93.98\%, HL of 0.018, and accuracy of 88.46\%. When comparing these results to the D5 dataset, which consists of 157 binary classes with 5 PCAPs per class, D20 shows a 3.73\% improvement in MAP, 3.63\% improvement in MAR, 41.94\% reduction in HL, and 12.89\% increase in accuracy. Similarly, when comparing D20 to D40, which has 17 binary classes and 40 PCAPs per class, D20 shows a 0.13\% reduction in MAP, 1.79\% improvement in MAR, 41.94\% reduction in HL, and 2.04\% increase in accuracy. D10 and D30 also show very similar comparison patterns with smaller margins. We can see that, in general, all metrics except HL do not show considerable changes. As a result, we can see that having more PCAPs (and, consequently, instances) is important for improving classifier performance, while the number of binaries does not seem to have a significant impact. However, when we observe the results of different datasets for the other three classifiers, we can see that it is the D5 dataset that shows the best and most consistent results, which we believe is the reason why \cite{DODIA2022} only presented the results for the D5 dataset\footnote{The results we obtained for D5 differ slightly from those presented in \cite{DODIA2022}. We were able to replicate the original results using their code, which manually splits the data samples into test and train sets in a deterministic manner. However, when we shuffled the dataset and randomly split the data into train and test sets, we obtained improved results, which are presented in this paper. We believe that the fixed nature of the original data split must have introduced some bias into the model.}.

\subsection{Discussion}

The above analysis addresses our first research question \textbf{(RQ1): Is it possible to improve the classification performance for identifying malware classes in Tor traffic?} The results undeniably provide a positive answer. However, it is crucial to consider the implications of these findings in a broader context.
As previously emphasised, accurate malware classification empowers organisations to develop more robust defenses against evolving malware threats. Intrusion detection systems produce a high volume of alerts, and even a minor improvement in detection accuracy can significantly enhance efficiency and minimise operational costs. Our experiments showcase a remarkable advancement in the detection performance for Tor-based malware, with both micro-average precision and recall rising from roughly 70\% to 90\%. Similarly, accuracy, which reflects the percentage of entirely correct predictions, has also witnessed a substantial increase, climbing from 40\% to 80\%. These significant improvements highlight the critical value of our findings.

\textbf{Justification for LaMP selection}: As previously mentioned, LaMP was chosen for this study following a thorough evaluation of existing multi-label classification methods and preliminary experiments. However, several additional factors solidify the selection of LaMP. First, LaMP incorporates label dependencies through the label interaction graph. This message-passing mechanism between labels allows the model to capture inherent relationships between them. Second, LaMP integrates labels into the feature learning process, optimising the feature representation specifically for classifying correlated labels. Finally, LaMP exhibits robustness to label imbalance, a characteristic empirically supported by our findings. This resilience can be largely attributed to its label message passing mechanism.

\textbf{Novelty:} This work presents several unique contributions. First, it represents the pioneering application of a deep learning technique specifically for the task of classifying Tor-based malware traffic into distinct malware classes. Second, it marks the first instance of a neural network approach being utilised for the multi-label classification of network traffic originating from the Tor network. While the LaMP model itself is not entirely novel, the specific adaptation employed in this study, alongside the hyperparameter configurations detailed in Table \ref{tab:hyperparameters}, constitutes a novel contribution. Finally, LaMP possesses additional functionalities beyond the scope of this study, including inherent interpretability via attention mechanisms, structure-agnostic learning capabilities, the ability to be parallelised, and the potential to learn from sparse data collections \cite{LANCHANTIN2020}. These functionalities hold significant promise for future endeavors in multi-label classification tasks involving the Tor network and its associated traffic.


\section{Interpreting Malware Class Identification Results}
\label{sec:rq2}
Having presented the results of our classification experiments in Section \ref{sec:rq1}, we now turn to our second research question: \textbf{\textit{RQ2: Are we able to see what factors led the models to identify a specific malware class?}}

\subsection{Impact of individual classes for the final result.}

To begin our interpretation of the results, we first investigate how individual classes have affected the overall performance of each classifier. To ensure consistency, we will use the D5 dataset for all comparisons across all four techniques.

\begin{table}[h]
  \centering
  \caption{Class-wise precision and recall for each classification technique using D5 dataset.}
  \tabcolsep=0.09cm
  \scalebox{0.9}{
  \begin{tabular}{l|c|c|c|c|c|c|c|c}
    \toprule
    \multirow{2}{*}{Class} & \multicolumn{2}{c|}{BR} & \multicolumn{2}{c|}{CC} & \multicolumn{2}{c|}{LP} & \multicolumn{2}{c}{LaMP}\\
    \cline{2-9}
    & P(\%) & R(\%) & P(\%) & R(\%) & P(\%) & R(\%) & P(\%) & R(\%) \\
    \midrule
    Backdoor & 0 & 0 & 0 & 0 & 100 & 9 & 100 & 100 \\
    Downloader & 81 & 87 & 80 & 87 & 81 & 91 & 91 & 96 \\
    Grayware & 72 & 92 & 71 & 92 & 75 & 90 & 91 & 99 \\
    Keylogger & 0 & 0 & 0 & 0 & 100 & 50 & 100 & 100 \\
    Miner & 73 & 64 & 73 & 68 & 68 & 87 & 92 & 92 \\
    Ransomware & 83 & 81 & 82 & 82 & 78 & 91 & 96 & 85 \\
    Spyware & 0 & 0 & 0 & 0 & 100 & 15 & 100 & 93 \\
    Unknown & 0 & 0 & 100 & 2 & 68 & 31 & 100 & 77 \\
    Virus & 100 & 6 & 100 & 12 & 100 & 35 & 100 & 94 \\
    Worm & 100 & 22 & 100 & 22 & 100 & 57 & 100 & 73 \\
    \hline
    Micro-Average & 76 & 76 & 76 & 77 & 77 & 83 & 92 & 94 \\
    \bottomrule
  \end{tabular}}
  \label{tab:rq2_classwise_results}
\end{table}

Table \ref{tab:rq2_classwise_results} presents the class-wise precision and recall for each classification technique using the D5 dataset. First, let us consider the Binary Relevance (BR) technique. We can observe three distinct patterns here.

\begin{enumerate}
    \item Pattern 1 (P1): The Backdoor, Keylogger, Spyware, and Unknown classes all have zero precision and recall. In other words, the model has not been able to correctly predict any instances of these classes. Further investigation of the individual predictions revealed that there were no predictions at all for these classes.
    \item Pattern 2 (P2): The model achieves 100\% precision for the Virus and Worm classes, but the recall values are very low ($<25\%$). This suggests that the model makes some predictions for these classes, and all of these predictions are correct. However, the model does not detect the majority of instances. 
    \item Pattern 3 (P3): The Downloader, Grayware, Ransomware, and Miner classes have more balanced precision and recall values (around 70\%-80\%), indicating that a considerable number of instances are correctly predicted for each class.
\end{enumerate}

Extending our focus to the other three classifiers, we observe that Classifier Chains (CC) exhibit similar patterns to BR for all classes except "Unknown", which shows P2. For Label Powerset (LP), no classes exhibit P1 and all classes that exhibit P1 for CC exhibit P2. However, the recall values for Keylogger and Worm increase substantially. Meanwhile, LaMP shows considerable improvements in all classes, with more balanced recall and precision values. For LaMP, two classes show 100\% precision and recall, and four classes show 100\% precision and $>70\%$ recall. Downloader, Grayware, Miner, and Ransomware exhibit a very consistent pattern (P3) across all classifiers, while the values are comparatively higher in LaMP.

A major insight into the above behaviour can be obtained by referring to Table \ref{tab:class_distribution} and Figure \ref{fig:label_cooccur}. These clearly show the number of instances available for each individual class in the D5 dataset and how these classes co-occur. We can clearly distinguish that the P3 pattern is observed for the classes with more instances, which indicates that the number of samples significantly affects classifier performance. However, interestingly, LaMP is able to classify the classes with very few samples (Keylogger, Virus, Spyware, and Backdoor), which demonstrates why it is a better candidate for our task. We conducted a second experiment using only the classes with more than 300 instances, and Table \ref{tab:rq2_classwise_filtered_results} shows the results we obtained. There, we can clearly see an improvement in the performance of the BR, CC, and LP techniques, especially in terms of MAR. Therefore, it is evident that one of the main reasons for the lower performance of the above classifiers is the lack of sufficient data samples. We can also see that LaMP's performance has not changed significantly, which confirms that it can perform well even with a small number of classifier instances for a specific class.

\begin{table}[h]
  \centering
  \caption{Class-wise precision and recall for Downloader, Grayware, Miner, and Ransomware in the D5 dataset.}
  \tabcolsep=0.09cm
  \scalebox{0.9}{
  \begin{tabular}{l|c|c|c|c|c|c|c|c}
    \toprule
    \multirow{2}{*}{Class} & \multicolumn{2}{c|}{BR} & \multicolumn{2}{c|}{CC} & \multicolumn{2}{c|}{LP} & \multicolumn{2}{c}{LaMP}\\
    \cline{2-9}
    & P(\%) & R(\%) & P(\%) & R(\%) & P(\%) & R(\%) & P(\%) & R(\%) \\
    \midrule
    Downloader & 86 & 95 & 86 & 94 & 89 & 94 & 96 & 98 \\
    Grayware & 72 & 93 & 73 & 93 & 78 & 91 & 90 & 94 \\
    Miner & 79 & 68 & 82 & 75 & 79 & 82 & 82 & 88 \\
    Ransomware & 95 & 82 & 94 & 78 & 94 & 87 & 94 & 96 \\
    \hline
    Micro-Average & 80 & 90 & 81 & 90 & 84 & 90 & 92 & 95 \\
    \bottomrule
  \end{tabular}}
  \label{tab:rq2_classwise_filtered_results}
\end{table}

\subsection{Using SHAP to interpret results}

Having established the impact of the number of samples on classifier performance, we now delve deeper into the underlying mechanisms of prediction. In this section, we will explain how we use SHAP, a popular XAI technique, to interpret classifier outcomes.

\subsubsection{Features Extraction and Selection}

In their work, Dodia et al. \cite{DODIA2022} use 215 features to represent each malware trace. Of these, 175 are connection-level features presented in \cite{HAYES2016}, while the remaining 40 are host-level features introduced in \cite{DODIA2022}. We have listed all these features in Table \ref{tab:list_of_features} in Appendix \ref{sec:appendix_A}, and we will use their relevant indices in this section when referring to them. While we will discuss certain important features in detail that are relevant to our findings in this section, we will not do so for all features. Interested readers can refer to the original papers \cite{HAYES2016, DODIA2022} for more information on these features.

Before using SHAP on the dataset, we removed all features with a value of zero for all samples. This resulted in the removal of 43 features (143-174, 176, 178, 206-214) from the dataset. We also removed all samples with the labels Backdoor, Keylogger, Spyware, Unknown, Virus, and Worm. This is because we were not getting any predictions for these labels (in three classifiers), as we mentioned previously.

\subsubsection{Shapley Additive exPlanations (SHAP)}
SHAP is a technique based on game theory that can be used to explain the output of a machine-learning model \cite{LUNDBERG2017SHAP} by identifying each individual feature's contribution to the model's predictions. 

The Shapley value for feature $i$ in the context of model $f$ can be calculated using the following equation:
\begin{equation}
\phi_i(f) = \sum_{S \subseteq N \setminus \{i\}} \frac{|S|!(|N|-|S|-1)!}{|N|!} [f(S \cup \{i\}) - f(S)]
\end{equation}
where, $\phi_i(f)$ is the Shapley value for feature $i$ in the context of model $f$. $S$ is a subset of features (excluding $i$) from the set of all features $N$. $|S|$ is the number of features in subset $S$ and $|N|$ is the total number of features in the model. $f(S \cup \{i\})$ is the model's prediction when considering the features in subset $S$ along with feature $i$, while $f(S)$ is the model's prediction when considering only the features in subset $S$.

The equation iterates over all possible subsets $S$ excluding $i$ ($S \subseteq N \setminus \{i\}$) and calculates the difference in model predictions when $i$ is included compared to when it's excluded from the feature set. For each iteration, the model considers all possible orders of adding features to subset $S$ ($|S|!$) and all the possible combinations of the features not in $S$ ((|N|-|S|-1)!) and use these values to calculate the weighted average of those differences. In summary, the Shapley value, $\phi_i(f)$, quantifies the marginal contribution of feature $i$ to the prediction. If the value is higher, it indicates greater importance, as features that consistently have a larger impact on predictions receive higher Shapley values. 

\begin{figure*}[htbp]
\centering
\begin{subfigure}[b]{0.24\textwidth}
    \centering
    \includegraphics[width=\linewidth, keepaspectratio]{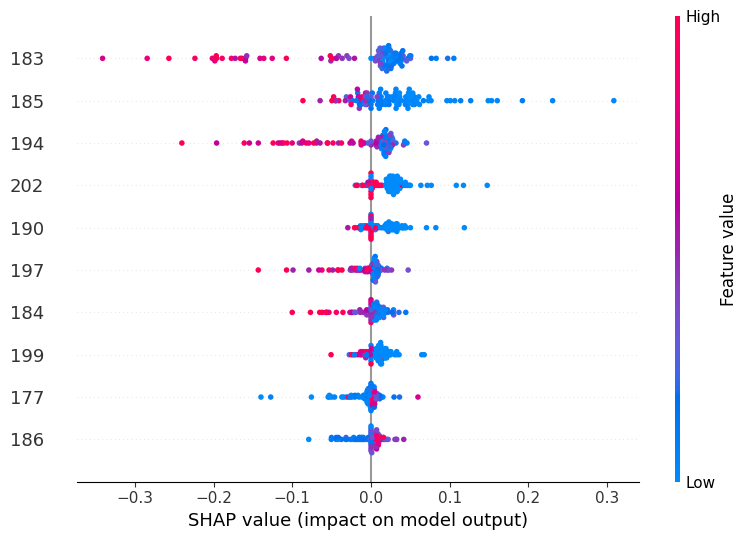}
    \caption{Downloader}
    \label{fig:br_downloader_summary}
\end{subfigure}
\begin{subfigure}[b]{0.24\textwidth}
    \centering
    \includegraphics[width=\linewidth, keepaspectratio]{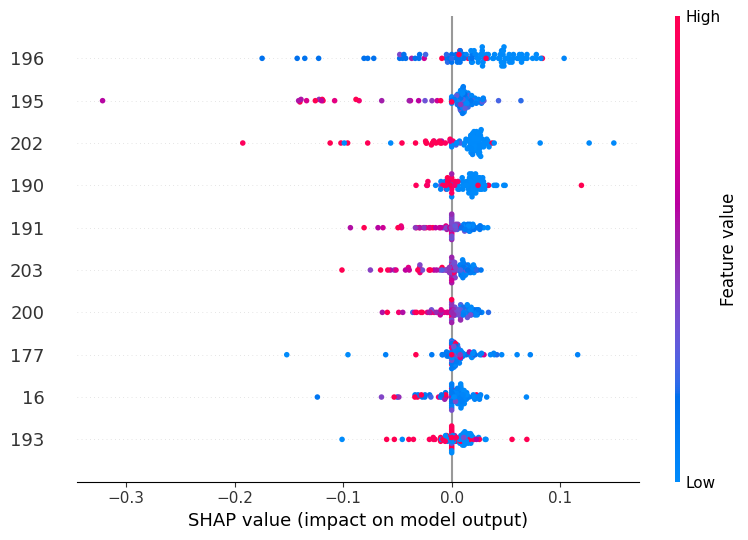}
    \caption{Grayware}
    \label{fig:br_grayware_summary}
\end{subfigure}
\begin{subfigure}[b]{0.24\textwidth}
    \centering
    \includegraphics[width=\linewidth, keepaspectratio]{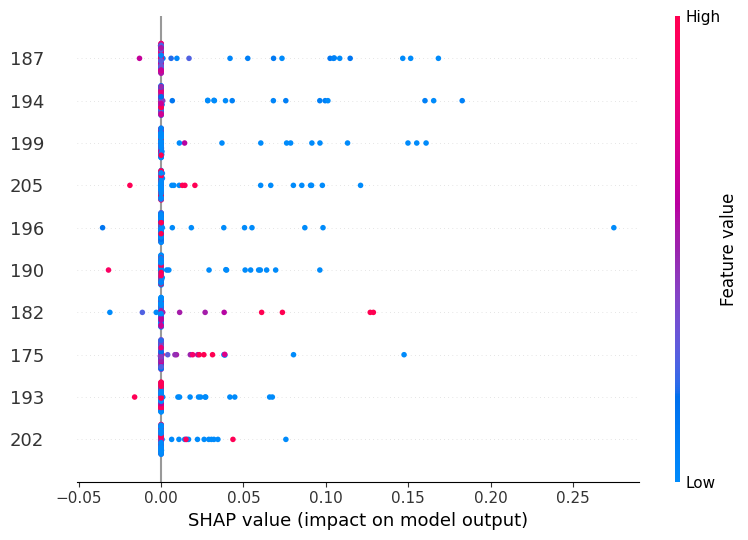}
    \caption{Miner}
    \label{fig:br_miner_summary}
\end{subfigure}
\begin{subfigure}[b]{0.24\textwidth}
    \centering
    \includegraphics[width=\linewidth, keepaspectratio]{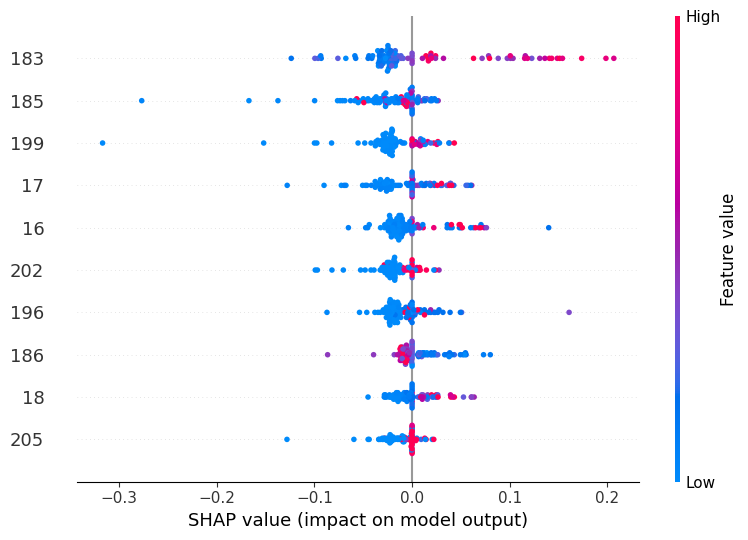}
    \caption{Ransomware}
    \label{fig:br_ransomware_summary}
\end{subfigure}
\caption{Summary plot for each class with the ten most important features - BR}
\label{fig:classwise_summary_plot}
\end{figure*}

\begin{figure}[h]
\centering
\includegraphics[width=0.6\linewidth]{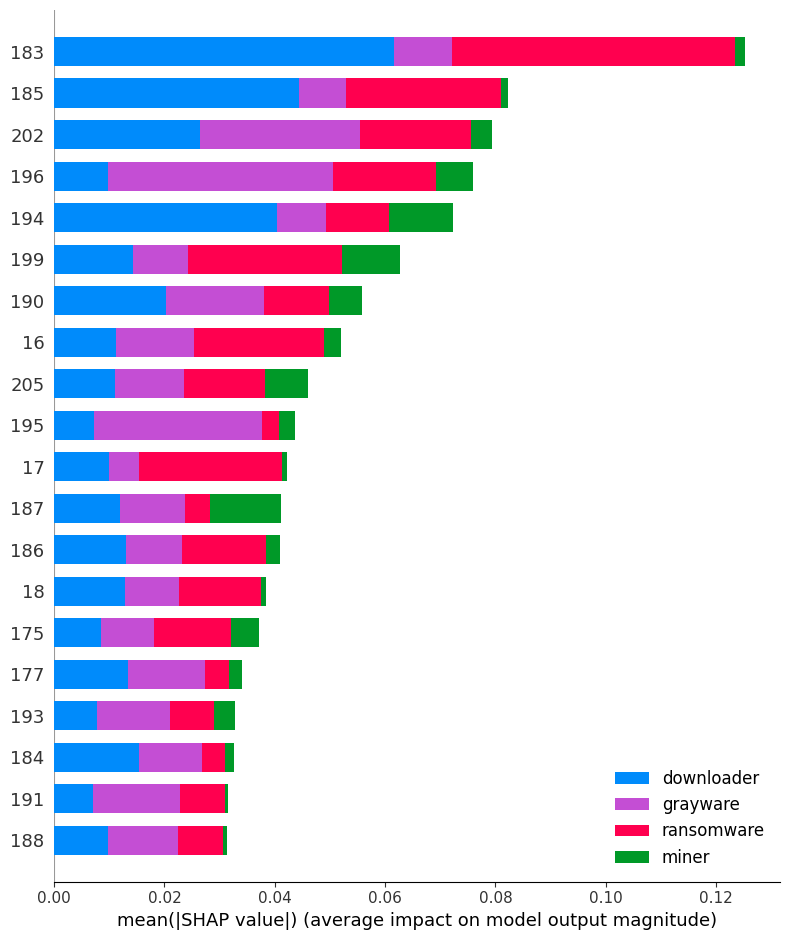}
\caption{Summary plot for feature importance - BR}
\label{fig:binary_relevance_summary}
\end{figure}

Now, let us consider the BR technique we have used in our paper and see how we can interpret the model predictions using SHAP. When interpreting a model, there are two main considerations: global interpretability and local interpretability. Global interpretability considers generic interpretations, such as which features have the greatest impact on the overall model output and how a certain feature can affect a specific class. Local interpretability considers an individual data sample and tries to interpret the prediction of that sample. We will focus mainly on the global intepretability in this paper. Local intepretability examples are discussed in Appendix \ref{sec:appendix_A} for an interested reader.

\subsection{Global Interpretability}
First, let us consider some global interpretations of the BR technique. Figure \ref{fig:binary_relevance_summary} shows the feature importance of the dataset D5 for each class. The y-axis represents the feature number (as specified in Table \ref{tab:list_of_features}), while the x-axis represents the average absolute Shapley values of a feature across the data. For better visualisation, we have only shown the 20 most important features.

Figure \ref{fig:binary_relevance_summary} reveals several interesting characteristics. First, most of the top 20 important features are host-level features, with only three connection-level features (based on percentiles of packet times) among them. Second, we can observe the relative importance of each feature for each class. For example, Feature 183, the average duration of Tor connections, has a significant impact on the Downloader and Ransomware classes, while Features 202 and 196 have a greater impact on Grayware than Feature 183. Similarly, Features 187, 194, and 199 have a larger impact on the Miner class than any of the previously mentioned features.

We can further gain insights into the predictive power of each feature by examining a summary plot\footnote{We can also use dependence plots to learn the impacts of a single feature on a model's outcome. More information can be found in Appendix \ref{sec:appendix_A}.} for each individual class. Figure \ref{fig:classwise_summary_plot} shows these individual summary plots for the Downloader, Grayware, Miner, and Ransomware classes. Similar to Figure \ref{fig:binary_relevance_summary}, the y-axis indicates the most important feature (from top to bottom) for each class, and the x-axis shows the SHAP values. A centre line at zero represents the model's expected output. Each point on the graph represents a data sample. Positive SHAP values contribute to improving the model's prediction for a given class, while negative SHAP values contribute to reducing it. The colour of each point represents the value of the feature. For example, in Figure \ref{fig:br_downloader_summary}, we can observe that high values of Feature 183 negatively affect the output, while low values positively affect the output. The longer the bar, the higher the impact of that feature on the prediction. Again, in Figure \ref{fig:br_downloader_summary}, we can clearly see that Feature 183 has a very strong impact on the model outcome for the Downloader class. Meanwhile, we can see an opposite pattern for the same feature in Figure \ref{fig:br_ransomware_summary}, the summary plot for Ransomware. This further explains the observations in Figure \ref{fig:binary_relevance_summary} on why Feature 183 has a huge impact on the overall model outcome. In summary, if the average duration of Tor connections is high, the trace is very likely to be Ransomware, while if it is low, there is a high likelihood that the trace is a Downloader. It is important to note that while these explanations help us to interpret the outcomes of the model, they may not necessarily support causation in the real world.

\subsection{Discussion}

\subsubsection{Extending the evaluation to all classifiers}

\begin{figure*}[ht]
\centering
\begin{subfigure}{0.3\linewidth}
  \centering
  \includegraphics[width=\linewidth, keepaspectratio]{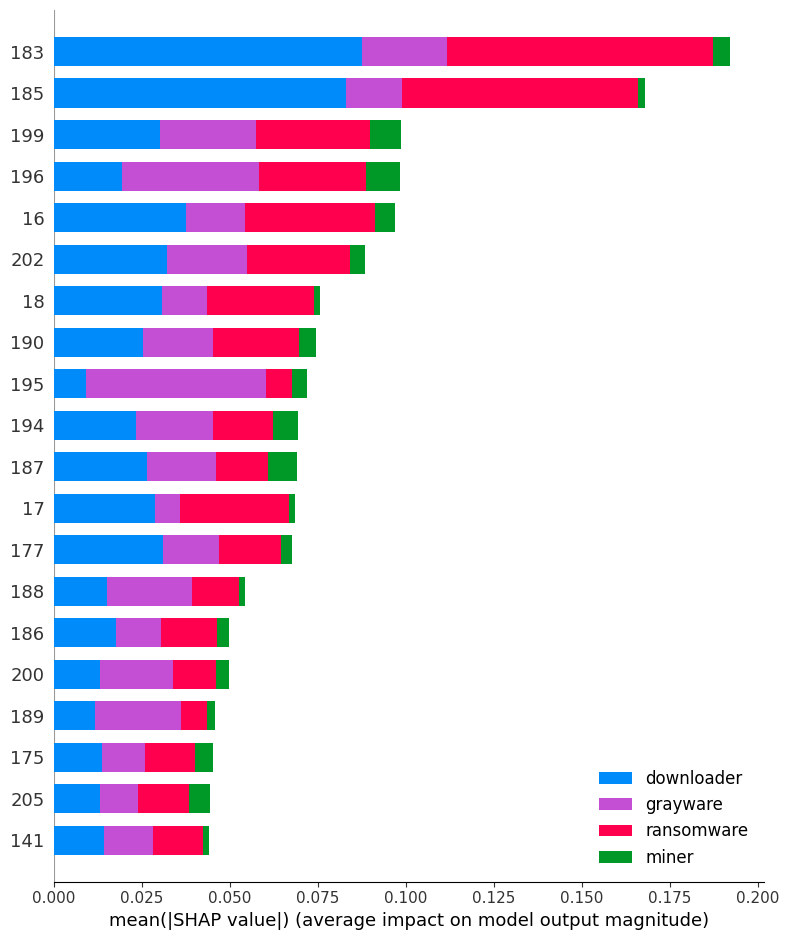}
  \caption{Classifier Chain}
  \label{fig:summary_plot_cc}
\end{subfigure}
\begin{subfigure}{0.3\linewidth}
  \centering
  \includegraphics[width=\linewidth, keepaspectratio]{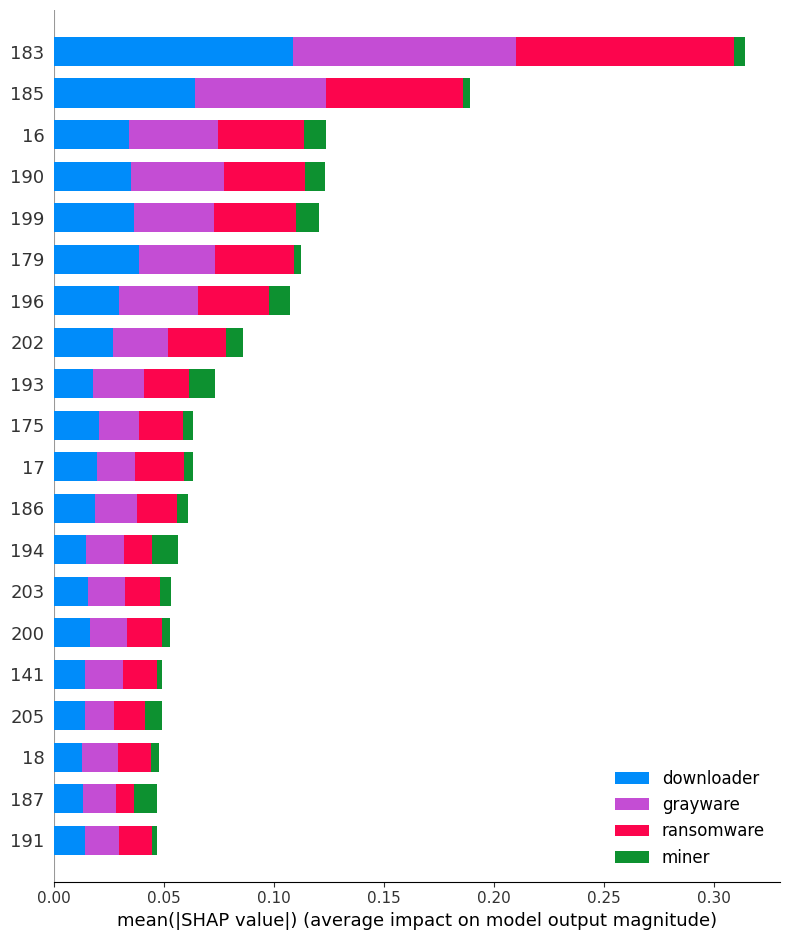}
  \caption{Label Powerset}
  \label{fig:summary_plot_lp}
\end{subfigure}
\begin{subfigure}{0.3\linewidth}
  \centering
  \includegraphics[width=\linewidth, keepaspectratio]{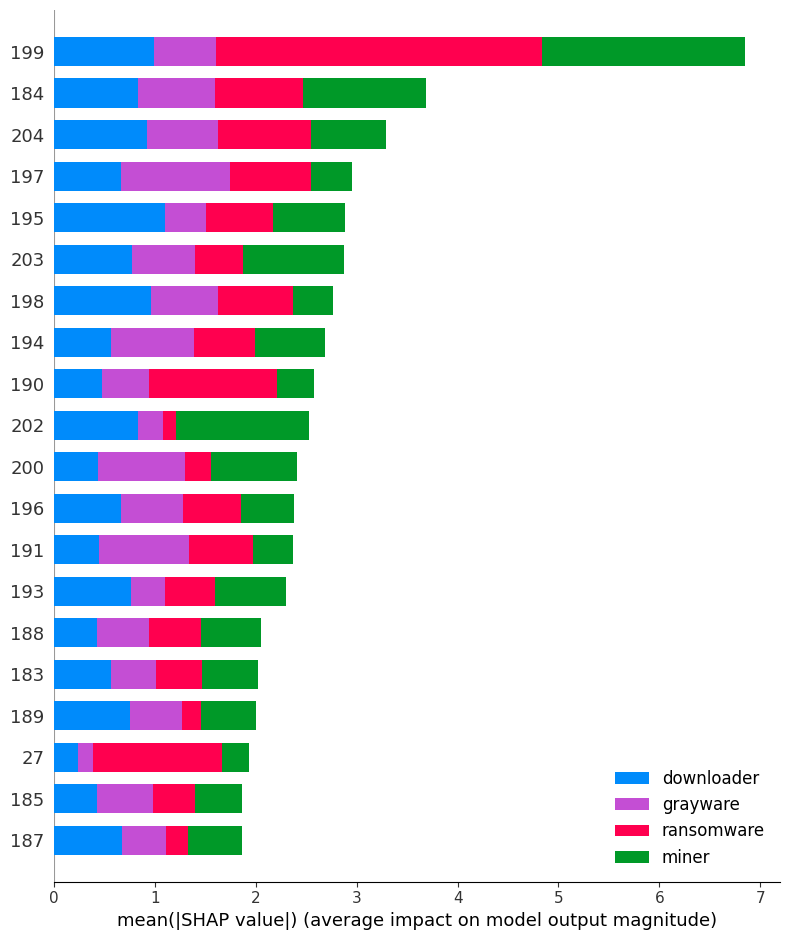}
  \caption{LaMP}
  \label{fig:summary_plot_lamp}
\end{subfigure}

\caption{Summary plot for feature importance - CC, LP, and LaMP}
\label{fig:summary_plots_cc_and_lp}
\end{figure*}

\begin{figure}[ht]
\centering
\includegraphics[width=0.6\linewidth]{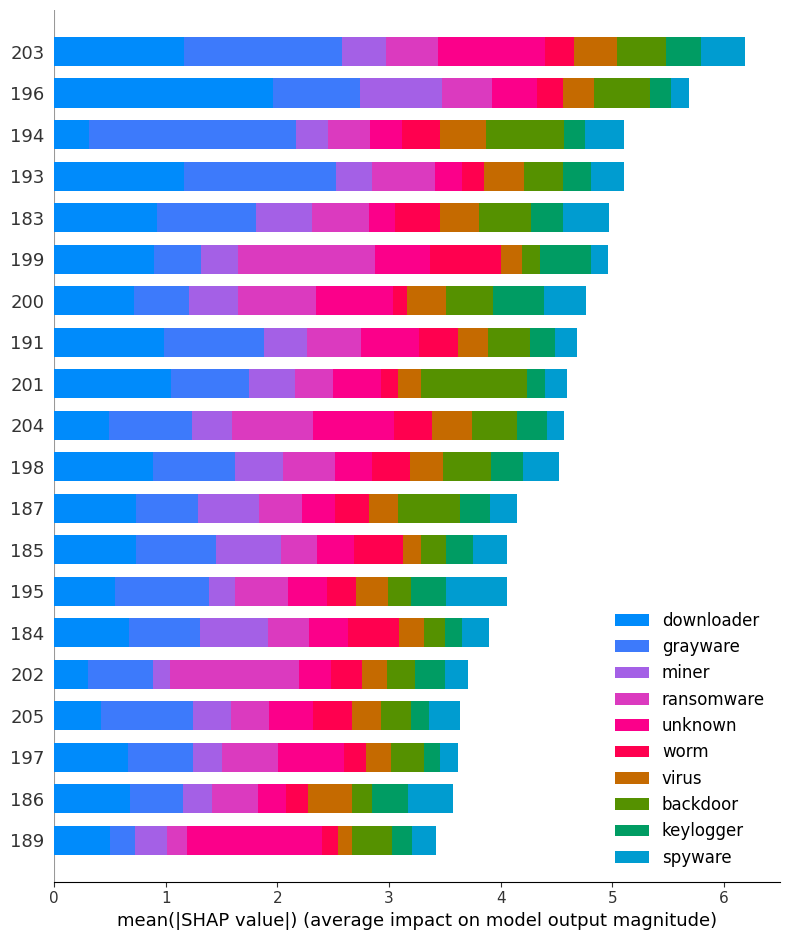}
\caption{Feature importance for all classes - LaMP}
\label{fig:lamp_summary}
\end{figure}

Having provided insights into global interpretability, we now proceed to examine and compare the most significant features across all classifiers utilised in this study. Figure \ref{fig:summary_plots_cc_and_lp} presents summary plots for CC, LP, and LaMP, revealing both similarities and differences in feature importance relative to the BR diagram shown in Figure \ref{fig:binary_relevance_summary}. Notably, Features 185 and 183 are prominent in the CC and LP predictions, as they are in the BR context. These features are also present in the LaMP plot, though their impact is less pronounced.

Additionally, Features 190, 202, and 199 are consistently important across all plots, indicating their model-agnostic nature, although the extent of their impact varies. A deeper insight into label dependencies can be obtained by comparing the BR plot with other plots. For instance, in the CC plot, Feature 185 shows a longer bar for Ransomware compared to BR, suggesting that the classification of Ransomware particularly benefits from label dependencies for this feature. In contrast, the LP and LaMP plots show nearly equal bars for Downloader, Grayware, and Ransomware, implying that the label sequence in classifier chains might not be optimal for Grayware and Miner classes. Notably, the LaMP plot demonstrates that Feature 185 has a uniform influence on the prediction of all classes when label dependencies are modeled through message passing, a trend that is consistently observed across all features with LaMP showing longer bars for the Miner class compared to other classifiers. This observation shows the unique ability of LaMP to leverage label dependencies effectively across different classes.

As BR does not account for label dependencies, comparing its plot with those of other classifiers can yield additional insights. For instance, if a feature exhibits bars of equal size in the BR plot and in plots of other models, it suggests that these features do not benefit significantly from considering label dependencies. Conversely, if the BR plot displays larger bars for all classes for certain features, while other models show smaller or varying bar sizes for these features, it may indicate that the BR model is over-relying on these features due to its assumption of label independence. This analysis helps in identifying the influence of label dependencies on feature importance across different classification models.

\subsubsection{Interpreting the results of LaMP for all classes}

When interpreting the results of LaMP, we can follow a similar approach to that used for the other classifiers. LaMP's summary plot for the main four classes (Figure \ref{fig:summary_plot_lamp}) shows some common traits with the summary plots of other classifiers, such as LaMP's primary reliance on host-level features for predictions. Additionally, we observed other interesting characteristics. For example, in the summary plots of other classifiers, the impacts of features for the "Miner" class were relatively lower compared to other classes. However, in LaMP, we could see that many of these features have a higher relative impact on identifying the "Miner" class. As LaMP was able to successfully identify all ten classes, we generated a summary plot for the entire dataset. Figure \ref{fig:lamp_summary} shows this plot, where we can see how the features impact all classes. Again, we can see that host-level features have a higher impact on the output. Interestingly, we can observe that the bars for the classes with a larger number of samples are longer than the bars for the classes with a smaller number of samples.

A slight concern when examining the SHAP values for LaMP is that they are much larger than the values obtained for the other classifiers. Directly comparing the range of these values across models can be misleading as the models have been trained differently and the SHAP values might have been normalised differently. However, we suspect that these very high values are a result of the actual inputs to the model, where a custom dictionary is used to map each feature value. To elaborate, assume a feature value ranges from 1 to 2, with individual sample values of 1.0, 1.1, 1.15, 1.23, and so on. A dictionary is created with each distinct value as the key and some integer as its corresponding value. This is done for each distinct value of every feature in the entire dataset. Finally, the feature values are mapped to the values in the dictionary\footnote{In the original LaMP paper \cite{LANCHANTIN2020}, the authors use several datasets with natural language inputs in their experiments by employing a similar mapping approach.}. For example, in the D5 dataset, there are 57,885 distinct original feature values. As a result, the mapped feature value may range from 1 to 57,885. This can lead to a wider range of SHAP values.

\subsubsection{Examining class-wise important features}

Now, let us discuss the top ten features of each class in more detail with respect to the BR classifier (Figure \ref{fig:classwise_summary_plot}). As mentioned previously, most of these features are host-level features. For the Downloader class, features based on the duration of Tor connections and the number of packets transmitted have a major impact on the prediction. Around five of the top ten impactful features for the Ransomware class are also in the list of top features for the Downloader class, but they show an opposite pattern, as previously explained with Feature 183. This implies that Ransomware and Downloader can be quite distinguishable from each other. Another interesting aspect of the top features of the Ransomware class is the presence of three connection-level features. According to Table \ref{tab:list_of_features}, these features are based on the percentiles of packet times. More specifically, they refer to the 25th, 50th, and 75th percentiles (1st, 2nd, and 3rd quartiles) of outgoing packet times of a given flow. It is quite surprising that out of 143 non-zero connection-level features, only three are among the top twenty important features. For Grayware, the most important features are the number of packets and the amount of data exchanged in Tor connections. Similar features also impact the predictions for the Miner class. The most unique observation in the Miner class is that it has very few negative points and a comparatively low number of positive points. Most of its values are on the centre line, which means they do not have much predictive power. Consequently, as shown in Table \ref{tab:rq2_classwise_filtered_results}, the BR model shows comparatively lower performance for the Miner class.

Further insights can be obtained by examining Figure \ref{fig:lamp_summary} - summary plot of the LaMP model. This figure highlights that Features 199, 185, and 184 significantly impact the classification of the Worm class, while Features 203, 194, and 186 are most influential for the Virus class. Feature 201 is predominantly influential in classifying Backdoor, and Features 199 and 200 are key for identifying Keylogger. Additionally, Feature 195 emerges as the most influential for Spyware classification.

Another important observation from this figure is the presence of considerably longer bars for the unknown class, a pattern similarly noted among the top four classes. This may be attributed to the relatively higher sample size of this class. This ability of LaMP to identify the unknown class is particularly valuable for anomaly detection tools, where it can help in recognising malicious traces without a known signature, thereby enhancing the robustness of security measures against emerging threats.

\subsubsection{Importance of this work}

As demonstrated throughout this section, \textbf{SHAP can be used to gain several insights into the outcomes of machine learning classifiers}, thus answering our second research question - RQ2. By looking at these different types of plots, we can better understand a model's performance. However, it is important to keep in mind that these interpretations are highly dependent on the models we are evaluating, and the models are dependent on the dataset and the features we provide. For example, the insights we obtained earlier indicating that the duration of a Tor connection has a positive relationship with a Ransomware trace and a negative relationship with a Downloader trace can be misleading if we didn't have any Ransomware traces with short durations and Downloader traces with long durations in our dataset. In any case, these observations are extremely important as they encourage us to look further into these relationships and try to understand important patterns rather than treating the machine learning model as a black box and accepting its outcomes.


\section{Evasion attack using adversarial perturbations}
\label{sec:rq3}
Having obtained several insights into how predictions are influenced by features in Section \ref{sec:rq2}, we attempted to exploit this information to trick the machine learning models. We began by using the BR technique as a case study and then extended our experiments to other classification techniques. Our goal was to investigate whether we could trick a trained model into making false predictions. We first discuss our threat model followed by the experimental approach, results, and analysis.

\subsection{Threat Model}

In the threat model outlined for this study, two primary entities are considered: a botmaster and a surveillance unit. The botmaster controls a C\&C server that utilises the Tor network for botnet and malware communications, embodying a malicious actor within the framework. Conversely, the surveillance unit, characterised as a resourceful and legitimate entity operating within legal parameters, is capable of monitoring and capturing network traffic. This unit not only detects that the traffic is associated with the Tor network but also possesses a trained machine learning model designed to identify various malware classes, with a particular focus on Ransomware as discussed in this paper.

For the following experiment conducted, it is assumed that the botmaster has configured the botnet specifically to deploy Ransomware. Moreover, it is presumed that the botmaster is aware of the Ransomware and Downloader samples that were used in the training of the surveillance unit’s machine learning model. This scenario sets the stage for a complex interplay between evasion tactics and detection capabilities, highlighting the challenges in cybersecurity threat modeling.

\subsection{Results and Analysis}

We used Ransomware samples as an example to investigate how the BR classifier predicts the Ransomware class. To do so, we employed the D5 dataset, which contains Downloader, Grayware, Miner, and Ransomware samples. We split the dataset 70:30 (train:test) as in the previous experiments and trained the model with the training dataset. However, in the testing dataset, we first filtered the samples so that only those with the Ransomware label remained, resulting in 54 samples. We then conducted three different experiments to try to answer our third research question: \textbf{\textit{RQ3: Is it possible to create variations of malware traffic (adversarial perturbations) that bypass our trained model and evade detection?}}

\textbf{Experiment 1 (E1):} In this experiment, we made no modifications to the original data samples. Instead, we used our trained model to generate predictions for them.

\textbf{Experiment 2 (E2):} In the second experiment, we sought to determine whether we could manipulate the model to predict Ransomware samples as Downloader samples. As shown in Figure \ref{fig:classwise_summary_plot}, Features 183 and 185 have a significant impact on both Ransomware and Downloader predictions, but in opposite directions. Recall that low values of Features 183 and 185 positively correlate with Downloader predictions. Therefore, in this experiment, we manually changed the values of these two features of the Ransomware-only test samples to the 25th percentile values of the Downloader-only test samples. The mean values of Feature 183 and 185 of the original Ransomware-only test samples were 21.14 and 48.01, respectively. We changed these values of all Ransomware-only test samples to 7.58 (183) and 13.63 (185). We then generated predictions for the modified samples with the same model used in E1.

\textbf{Experiment 3 (E3):} In this experiment, we aimed to determine whether we could prevent the model from identifying a Ransomware sample. To do so, we performed a perturbation similar to the one we performed in E2. However, instead of considering the top two features, we considered the top five features impacting the Ransomware outcome. We calculated the 10th percentile of each feature in Ransomware-only test samples and replaced the corresponding feature values of all samples with those 10th percentile values. As shown in Figure \ref{fig:br_ransomware_summary}, all of these top five features (183, 185, 199, 17, and 16) have a negative impact on the ransomware prediction if the values are low.

\begin{table}[h]
  \centering
  \caption{Predictions of the learned model for the 54 Ransomware only samples}
  \tabcolsep=0.09cm
  \scalebox{0.9}{
  \begin{tabular}{c|c|c|c|c|c|c|c|c|c|c|c|c}
    \toprule
    \multirow{2}{*}{Class} & \multicolumn{3}{c|}{BR} & \multicolumn{3}{c|}{CC} & \multicolumn{3}{c|}{LP} & \multicolumn{3}{c}{LaMP}\\
    \cline{2-13}
    & E1 & E2 & E3 & E1 & E2 & E3 & E1 & E2 & E3 & E1 & E2 & E3\\
    \midrule
    Downloader & 7 & 33 & 39 & 7 & 32 & 38 & 5 & 33 & 43 & 3 & 7 & 8\\
    Grayware & 16 & 28 & 36 & 17 & 28 & 31 & 7 & 33 & 45 & 6 & 8 & 10\\
    Miner & 0 & 0 & 0 & 0 & 0 & 0 & 0 & 0 & 0 & 0 & 0 & 0\\
    Ransomware & 44 & 37 & 7 & 43 & 22 & 12 & 48 & 21 & 8 & 46 & 45 & 45\\
    \bottomrule
  \end{tabular}}
  \label{tab:evasion_attack_results}
\end{table}

Table \ref{tab:evasion_attack_results} shows the predictions by all classifiers for Downloader, Grayware, Miner, and Ransomware classes in Experiments E1, E2, and E3. Note that this is a multi-label classification task, so the sum of the predictions can be greater than the number of samples (54 in this case). For E1, all classifiers were able to predict a considerable number of samples as Ransomware. There were a few samples that were misclassified as Downloader. In E2, the number of Ransomware predictions by each classifier decreased while the Downloader predictions increased. However, we had only modified the values of two of the features used. This result shows that we can trick the learned models into classifying a traffic trace incorrectly with simple adversarial perturbations. The results for E3 further confirm this finding, as we were able to trick the models into not identifying a considerable number of Ransomware samples. 

\textbf{Increase in Grayware predictions:} We can also notice that the minor changes we made caused an increase in the number of Grayware predictions. This is more concerning as Grayware is usually considered less harmful than both Ransomware and Downloader. Therefore, if a Ransomware sample can easily be modified to be classified as Grayware, it can avoid many malware detection and prevention systems and cause more infections and damage. As for the reasons behind this characteristic, there exists training data containing several samples that have label combinations of Ransomware-Downloader and Ransomware-Grayware-Downloader (see Figure \ref{fig:label_cooccur}). It is possible that when we modify the above features to reflect more Downloader characteristics, classifiers can identify them as having multiple labels, including Grayware.

\textbf{Robustness of LaMP:} However, we can observe that LaMP is the only model that showed very little deviation for E2 and E3, where its Ransomware predictions were only reduced by one, while other predictions (Downloader and Grayware) were only slightly increased. If we consider Figure \ref{fig:summary_plot_lamp}, we can observe that the features we modify are not the most important features when it comes to the outputs of LaMP. Even though we used the same set of features for better comparison, it is possible that LaMP is using a different set of features for classifying these samples. In any case, we have still been able to deceive LaMP into giving a few false Downloader and Grayware predictions, which means that even LaMP is not 100\% robust. To gain further insights into these predictions, we used the decision plots we used in Section \ref{sec:rq2}. The decision plots for the Downloader and Ransomware classes in the above experiments (presented in Figure \ref{fig:br_evasion_attack_decision_plot} in Appendix \ref{sec:appendix_A}) clearly demonstrated how the changes in these features affect the final predictions.

\subsection{Discussion}

In summary, this experiment provides us with two major insights. Firstly, \textbf{it demonstrates the vulnerability of high-performing classifiers to adversarial manipulation despite their apparent accuracy in identifying malware classes}. As noted earlier, current traffic analysis methodologies often neglect robust evaluation of their models' resilience to attack. In this specific context of classifying malware traces within Tor traffic, LaMP's dual strength in accuracy and robustness establishes it as the superior model. Secondly, the experiment strengthens the case for interpretable machine learning models. By simply understanding the decision-making processes behind class predictions and identifying the most crucial features, we successfully constructed a simple yet effective evasion attack. This attack's effectiveness further corroborates the accuracy of the interpretations outlined in the previous section.

While the above attack successfully deceived the machine learning models in our simulated experiments E2 and E3, it remains crucial to assess the practical feasibility of such attacks in a real-world setting. This necessitates exploring the potential for modifying the relevant feature values under realistic conditions. To recall, the five features are as follows.

\begin{itemize}
    \item 183: Average duration of Tor connections (seconds)
    \item 185: Max duration Tor connection per PCAP/host - seconds
    \item 199: Mode number of packets sent in all Tor connections
    \item 17: 50th percentile of the list of outgoing packets (in order) of a given flow
    \item 16: 25th percentile of the list of outgoing packets (in order) of a given flow
\end{itemize}

These five features depend on three parameters: the duration of a Tor connection, the number of packets in a Tor connection, and the outgoing packets in a Tor connection. In our attack scenario, we aim to reduce the values of these features.

\textbf{Reducing the connection duration:} Malware creators can configure C\&C servers to create short-lived Tor connections by establishing and terminating the connections quickly, ensuring that the average connection time is low. Additionally, they can implement connection limits and session timeouts (to disconnect from Tor) to ensure that the maximum duration does not exceed a certain threshold.

\textbf{Reducing the number of packets sent:} C\&C servers can be configured to send a low number of packets in each Tor connection and to slow down the packet transmission rate in conjunction with short-lived connections. This will also reduce the number of packets sent in a Tor connection, thereby affecting the mode.

\textbf{Reducing the percentile values:} Instead of transmitting large chunks of data, an adversary can configure the C\&C server to split it into smaller chunks and transmit them in separate connections. Additionally, transmitting varying amounts of data in different connections can introduce greater fluctuations in the aforementioned values, making it more difficult for a machine learning model to accurately identify a malware trace.

As explained above, malware creators can easily make the above changes to their malware communications and transmissions. Therefore, we can conclude that the models are highly susceptible to this type of evasion attack in practice.


\section{CONCLUSION and FUTURE WORK}
The rapid growth of malware and botnets leveraging the Tor network poses substantial difficulties due to their intrinsic anonymity. Furthermore, the limited availability and difficulty in generating comprehensive datasets impede further research efforts in this domain. While prior studies have achieved some success in identifying malware concealed within Tor traffic, precisely classifying the various malware classes continues to be a significant challenge. This investigation aims to bridge this knowledge gap by exploring three critical research questions.

First, we evaluated various machine learning techniques for malware class identification. Among them, LaMP, an MPNN-based technique, achieved superior performance, exceeding 90\% micro-average precision and recall in multi-label classification. Utilising the dataset from \cite{DODIA2022}, one of the most extensive in this area, LaMP demonstrated a significant performance improvement compared to previously explored methods like BR, CC, and LP. For instance, on the D20 dataset, LaMP achieved a 27\% improvement in MAP, 22.08\% in MAR, 78.05\% in HL, and 65.69\% in accuracy compared to LP, the next best performer.

Motivated by the limitations of "black-box" machine learning models in the literature, our second research question focused on interpreting the results of four classifiers using SHAP, a prominent XAI technique. Through SHAP-generated values and visualisations: summary plots, decision plots, force plots, and dependence plots; we analysed feature importance at both global and individual sample levels.
Our analysis of feature importance provided insights into the factors influencing class predictions. By exploiting these insights, the final research question explored the classifiers' susceptibility to adversarial attacks. Manual manipulation of high-impact features enabled the generation of false positives and negatives, demonstrating their limitations in robustness. For instance, we observed that altering a Tor connection's duration and packet count could trick a classifier into misclassifying a Ransomware sample as the less severe Grayware. This underscores the significance of model interpretability and necessitates further investigation into developing robust algorithms for Tor traffic analysis.

This study is primarily limited by the dependence on the AVClass tool \cite{AVCLASS2016, AVCLASS22020} for generating the labelled dataset. 
Therefore, the accuracy of our findings hinges on the underlying tool's accuracy. While the generalisation capabilities of the employed machine learning models exhibit promise, future endeavors should prioritise dataset refinement. This encompasses qualitative enhancements, such as guaranteeing accurate labelling and a representation of malware types—and quantitative expansion — e.g., increasing the number of samples for specific classes  like Virus, keylogger, Spyware, and Worm. Furthermore, subsequent research could investigate the applicability of these models in diverse contexts, including the analysis of varied botnet activities.
\label{sec:conclusion}

\bibliographystyle{ieeetr}
\bibliography{main}

\appendix
\section{Appendix A}
\label{sec:appendix_A}
\subsection{Evaluation Metrics}
\subsubsection{Micro-Average Precision (MAP)}
MAP is calculated by summing up the true positives for all classes and dividing it by the sum of true positives and false positives for all classes. It gives equal weight to each instance regardless of its class.
\begin{equation}
    MAP = \frac{\sum_{i=1}^{N} {TP}_i}{\sum_{i=1}^{N} {TP}_i + \sum_{i=1}^{N} {FP}_i},
\end{equation}
where $N$ is the total number of classes, ${TP}_i$ is the number of true positives for class $i$, and ${FP}_i$ is the number of false positives for class $i$.

\subsubsection{Micro-Average Recall (MAR)} 
MAR is calculated by summing up the true positives for all classes and dividing it by the sum of true positives and false negatives for all classes.
\begin{equation}
    MAR = \frac{\sum_{i=1}^{N} {TP}_i}{\sum_{i=1}^{N} {TP}_i + \sum_{i=1}^{N} {FN}_i},
\end{equation}
where ${FN}_i$ is the number of false negatives for class $i$. $N$ and ${TP}_i$ are same as above.

\subsubsection{Accuracy (AC)}
Accuracy represents the proportion of correctly classified instances out of the total number of instances in the dataset.
\begin{equation}
    Accuracy = \frac{TP + TN}{TP + TN + FP + FN},
\end{equation}
where $TP$, $TN$, $FP$, and $FN$ represent true positives, true negatives, false positives and false negatives, respectively.

\subsubsection{Hamming Loss (HL)}
Hamming Loss measures the fraction of labels that are incorrectly predicted to the total number of labels. It can be calculated as follows.
\begin{equation}
    HL = \frac{1}{NL} \sum_{i=1}^{N} \sum_{j=1}^{L} (P_{ij} \oplus A_{ij}),
\end{equation}
where $N$ is the total number of instances in the dataset, $L$ is the total number of labels, $P_{ij}$ represents whether the $j^{th}$ label for the $i^{th}$ instance is predicted correctly (1 if correct, 0 if incorrect), $A_{ij}$ represents whether the $j^{th}$ label for the $i^{th}$ instance is an actual label (1 if true, 0 if false), and $\oplus$ is the XOR symbol, which returns 1 if only one of the operands is 1, but not both. 

\subsection{List of Features}

All the features used in our experiments are listed in Table \ref{tab:list_of_features}. The order in which they are listed (their index numbers) is used in our Figures and analyses.

\begin{table*}[h]
  \centering
  \caption{List of Features used in this work.}
  \tabcolsep=0.09cm
  \scalebox{0.8}{
  \begin{tabular}{c|p{8cm}|c|p{9cm}}
    \toprule
    Index & Feature & Index & Feature\\
    \midrule 
    0 - 11 & Statistical features (mean, max, std. deviation, etc.) based on Inter-arrival times & 176 & Number of failed/rejected Tor connections attempts: connection\_state = S0/REJ\\
    12 - 23 & Features based on percentiles of packet times & 177 & Rate of Tor connections: Average number of Tor connections per second\\
    24 - 26 & Features based on the number of packets & 178 & Rate of failed attempts: Number of failed attempts/second\\
    27 - 30 & Features based on the First and last 30 packets of a flow & 179 & Number of unique destination ports used across Tor connections\\
    31 & Feature based on packet concentration (std. deviation) & 180 & Most used destination port\\
    32 & Feature based on packet concentration (mean) & 181 & Number of non-standard destination ports seen in Tor connections\\
    33 & Feature based on number of packets per second (mean) & 182 & Most frequent non-standard destination port used\\
    34 & Feature based on number of packets per second (std. deviation) & 183 & Average duration of Tor connections (seconds)\\
    35 & Feature based on outgoing packet ordering (mean) & 184 & Smallest duration Tor connection seen per host/PCAP\\
    36 & Feature based on incoming packet ordering (mean) & 185 & Max duration Tor connection per PCAP/host - seconds\\
    37 & Feature based on outgoing packet ordering (std. deviation) & 186 & Number of short-duration Tor connections (max 1 minute)\\
    38 & Feature based on incoming packet ordering (std. deviation) & 187 & Average time gap between each Tor connection\\
    39 & Feature based on packet concentration (median) & 188 - 190 & Mean/median/mode of the total transmitted packets in a Tor connection\\
    40 & Feature based on number of packets per second (median) & 191 - 193 & Mean/median/mode of the total data exchanged in Tor connections\\
    41 & Feature based on number of packets per second (min) & 194 - 196 & Mean/median/mode of the total data sent in all Tor connections\\
    42 & Feature based on number of packets per second (max) & 197 - 199 & Mean/median/mode number of packets sent in all Tor connections\\
    43 & Feature based on packet concentration (max) & 200 - 202 & Mean/median/mode number of packets received in all Tor connections\\
    44 & Percentage incoming packets & 203 - 205 & Mean/median/mode of the total data received in all Tor connections\\
    45 & Percentage outgoing packets & 206 & Total number of DNS queries rcode\_name: rcode:599: NXDOMAINS\\
    46 - 116 &  Alternative concentration features & 207 & Total number of DNS queries rcode\_name: rcode:596: REFUSED\\
    117 - 137 &  Alternative number of packets per second features & 208 & Total number of DNS queries rcode\_name: rcode: 2 : SERVFAIL\\
    138 & Sum of alternative concentration features & 209 & Total number of onion domain accesses\\
    139 & Sum of alternative number of packets per second features & 210 & Total number of unique onion domains accessed (seen in DNS logs)\\
    140 & Sum of inter-arrival time based features & 211 & Total number of 'rejected' onion domain queries\\
    141 & Sum of packet time percentile features & 212 & Total number of onion domains accessed\\
    142 & Sum of number of packet-based features & 213 & Total number of links with 'consensus' seen \\
    143 - 174 & Padded features (with zeros) & 214 & Total number of URLs with "tor" keyword  \\
    175 & Number of Tor connections\\
    \bottomrule
  \end{tabular}}
  \label{tab:list_of_features}
\end{table*}

\subsection{Local Interpretability}

\begin{figure*}[h]
\centering
\begin{subfigure}{\linewidth}
  \centering
  \includegraphics[width=0.7\linewidth, keepaspectratio]{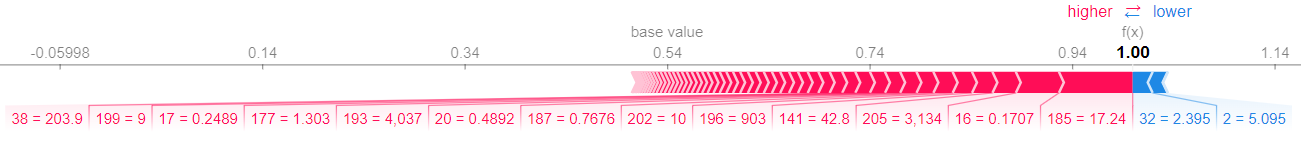}
  \caption{Downloader}
  \label{fig:sample_15_downloader}
\end{subfigure}
\begin{subfigure}{\linewidth}
  \centering
  \includegraphics[width=0.7\linewidth, keepaspectratio]{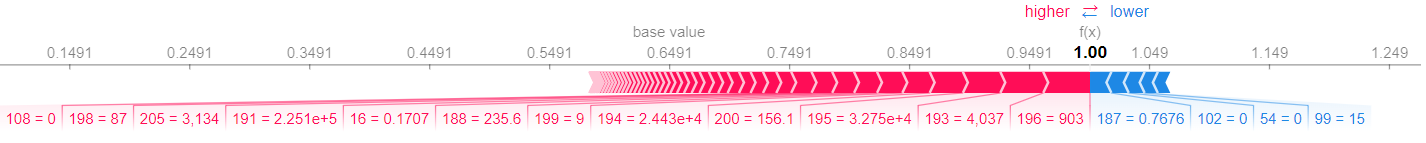}
  \caption{Grayware}
  \label{fig:sample_15_graware}
\end{subfigure}
\begin{subfigure}{\linewidth}
  \centering
  \includegraphics[width=0.7\linewidth, keepaspectratio]{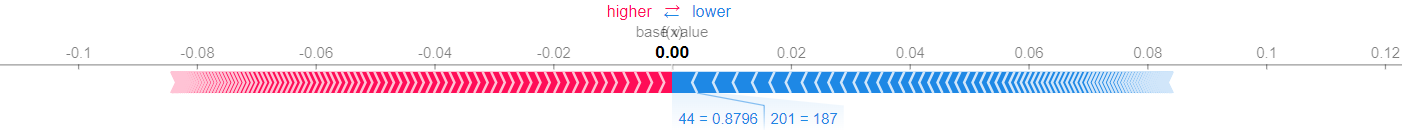}
  \caption{Miner}
  \label{fig:sample_15_miner}
\end{subfigure}
\begin{subfigure}{\linewidth}
  \centering
  \includegraphics[width=0.7\linewidth, keepaspectratio]{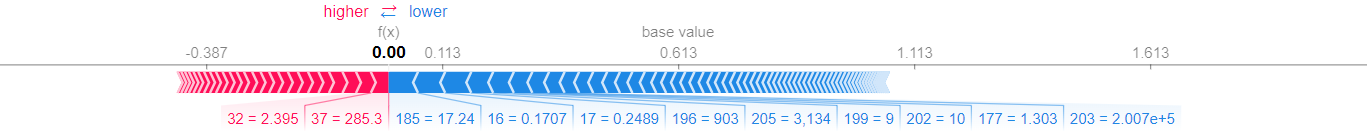}
  \caption{Ransomware}
  \label{fig:sample_15_ransomware}
\end{subfigure}
\caption{Force plots for predicting Downloader, Grayware, Miner, and Ransomware classes for a sample with the actual labels \textit{Downloader} and \textit{Grayware} - BR}
\label{fig:sample_15_force_plot}
\end{figure*}

Previously, in Section \ref{sec:rq2}, we interpreted the model outcomes in a more general manner, considering many data points at once. Here, we use SHAP to examine the individual predictions of a sample. Figure \ref{fig:sample_15_force_plot} shows individual force plots for predicting Downloader, Grayware, Miner, and Ransomware classes by the BR classifier for a data sample with the actual labels "Downloader" and "Grayware." First, consider one of the subplots (Figure \ref{fig:sample_15_downloader}) and attempt to interpret the results. This diagram shows a base value of 0.54 on the top line, which is the reference point (baseline) used by the model to make a prediction for the Downloader class. The length of the bar and the direction of the arrowhead for each feature represent the feature's contribution and the direction of influence on the prediction, respectively. For example, Features 185, 16, 205, and 141 positively influenced the prediction, while Features 32 and 2 negatively influenced it. The cumulative contributions from the features and the baseline provide the final prediction for the class ($f(x) = 1.00$). It means that the model identifies the sample as a Downloader. Similarly, if we consider Figure \ref{fig:sample_15_graware}, \ref{fig:sample_15_miner}, and \ref{fig:sample_15_ransomware}, we can observe $f(x) = 1.00$, $f(x) = 0.00$, and $f(x) = 0.00$, respectively, which means that the model predicts Grayware and does not predict Miner or Ransomware as labels.

\begin{figure}[h]
\centering
\includegraphics[width=0.75\linewidth]{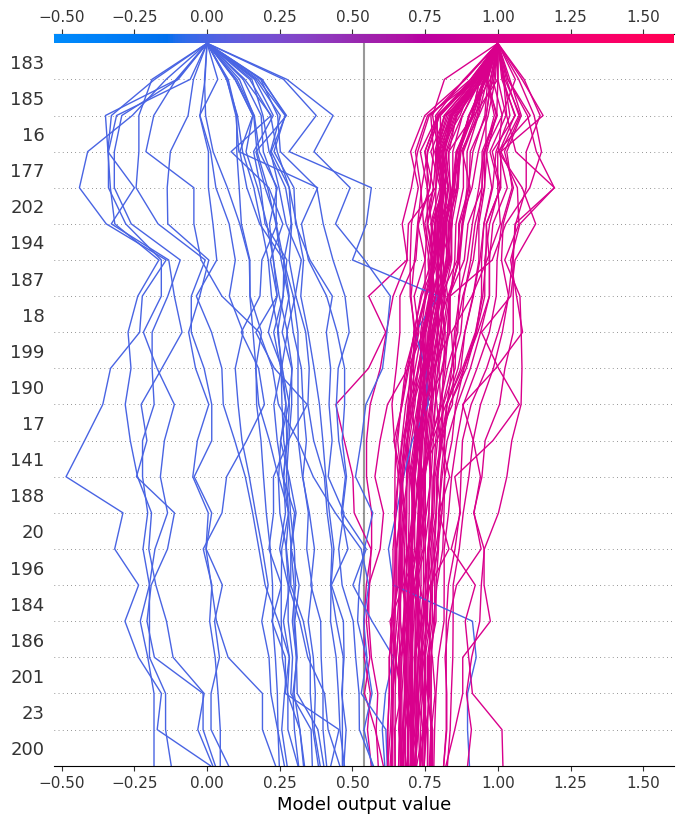}
\caption{Decision plot for Downloader class - BR}
\label{fig:br_decision_downloader}
\end{figure}

Next, we will examine another SHAP-related plot type: the Decision plot for the Downloader class prediction (Figure \ref{fig:br_decision_downloader}). It shows how the classifier predicted multiple data samples. The horizontal axis at the bottom shows the baseline output of the classifier affected by all the features except for the ones shown in the diagram. (The baseline of a model for a specific class is the same for all data samples in the same dataset.) The horizontal line at the top shows the final prediction, which is whether the sample is a Downloader or not. The vertical axis shows the top features affecting the final prediction and indicates their level of importance. In this plot, we can see that a small subset of features is strong enough to completely reverse the impact of all the other features.

\subsection{Additional plots}
\begin{figure*}[h]
\centering
\begin{subfigure}{0.35\linewidth}
  \centering
  \includegraphics[width=\linewidth, keepaspectratio]{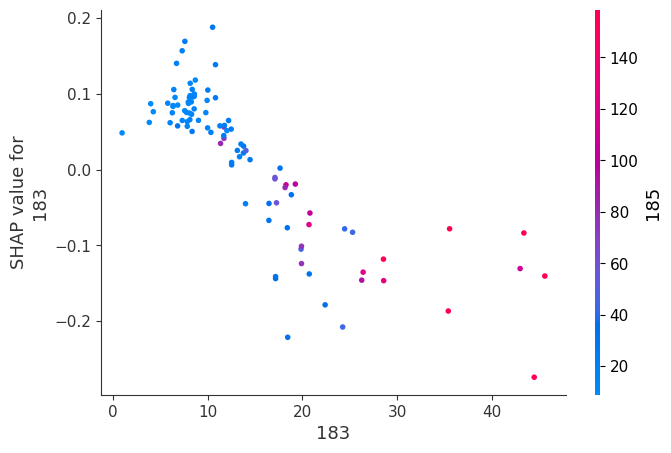}
  \caption{Downloader}
  \label{fig:br_downloader_interaction}
\end{subfigure}
\begin{subfigure}{0.35\linewidth}
  \centering
  \includegraphics[width=\linewidth, keepaspectratio]{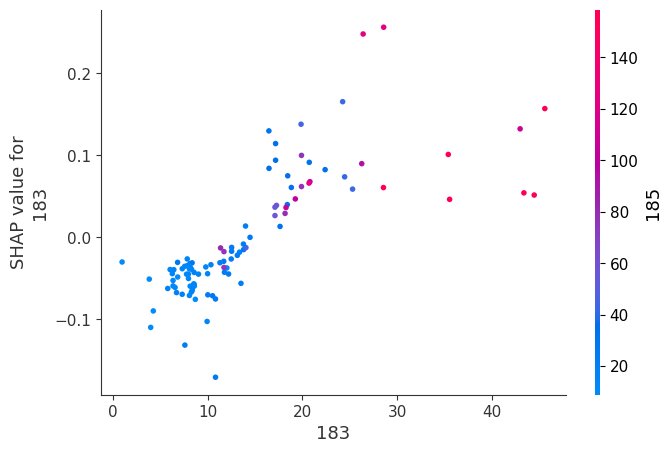}
  \caption{Ransomware}
  \label{fig:br_ransomware_interaction}
\end{subfigure}

\caption{Dependence plot for average duration of Tor connections with maximum duration Tor connection per PCAP/host (seconds) - BR}
\label{fig:183_175_feature_interaction}
\end{figure*}

In addition to the summary plots discussed in Section \ref{sec:rq2}, we also used dependence plots to gain insights into the model's functionality (from a global interpretation perspective). We generated a sample plot for Feature 183, with Feature 185 as the second feature, as shown in Figure \ref{fig:183_175_feature_interaction}. Since we cannot show all feature pairs for all classes, here, we only show the top two features for the Downloader and Ransomware classes to demonstrate the use of these plots. In dependence plots, the x-axis shows the values of the primary feature we are investigating, and the y-axis represents the SHAP values for that feature. The colour gradient represents the magnitude of the second feature. In Figure \ref{fig:br_downloader_interaction}, we can see that when the average duration of a Tor connection is low, the SHAP values are quite high, indicating a positive impact on the prediction. However, as the values increase, the SHAP value decreases, showing an inverse relationship between the feature and the model outcome. Additionally, when the average duration of the Tor connection increases, we can also notice that the maximum duration of the Tor connection also increases (via the colour coding). Therefore, we can conclude that these two features have a similar impact on the model's prediction. When we consider Figure \ref{fig:br_ransomware_interaction}, we can see that when the duration increases, it positively affects the model prediction. 

\begin{figure*}[h!]
\centering
\begin{subfigure}[b]{0.25\textwidth}
    \centering
    \includegraphics[width=\linewidth, keepaspectratio]{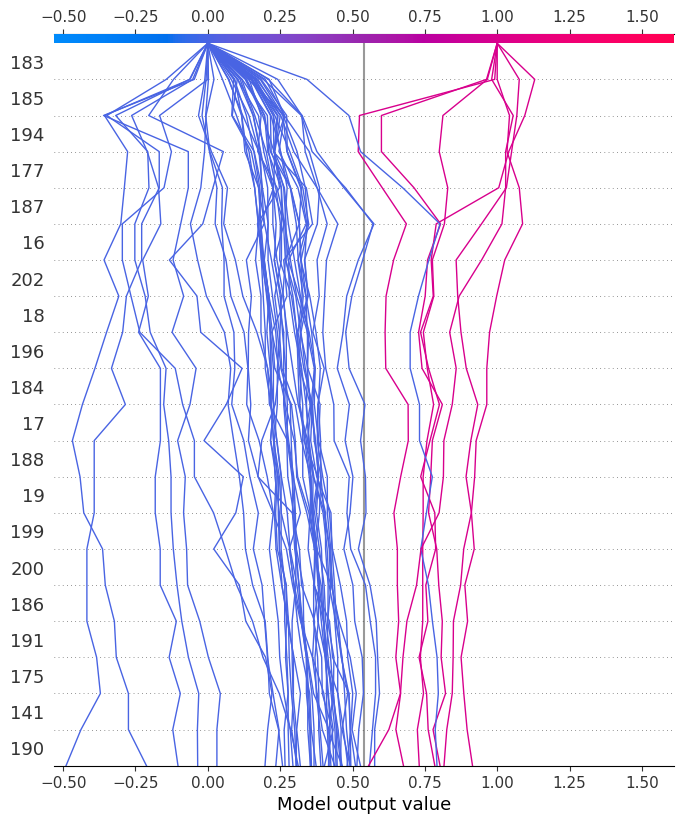}
    \caption{E1: Downloader}
    \label{fig:br_evasion_attack_ex1_downloader}
\end{subfigure}
\begin{subfigure}[b]{0.25\textwidth}
    \centering
    \includegraphics[width=\linewidth, keepaspectratio]{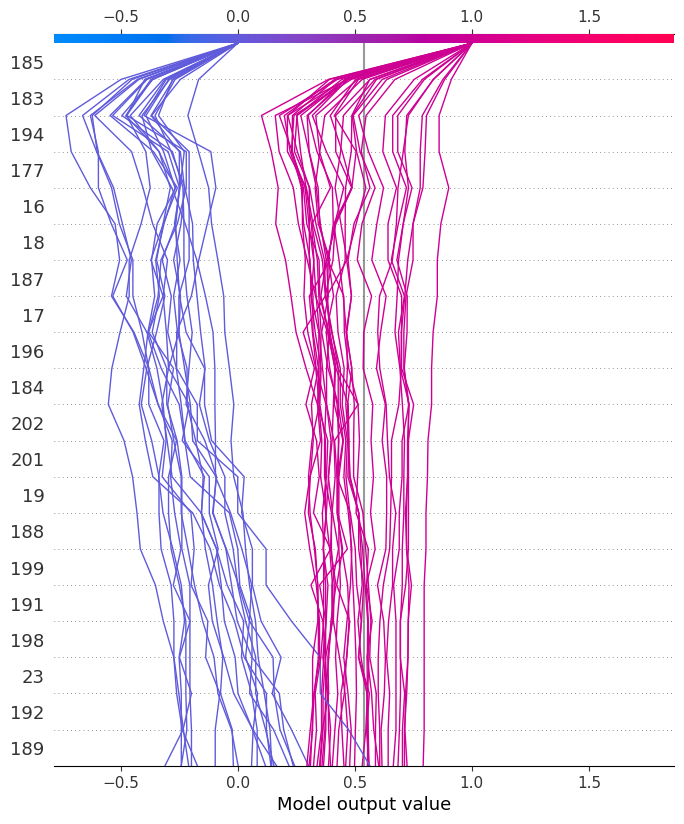}
    \caption{E2: Downloader}
    \label{fig:br_evasion_attack_ex2_downloader}
\end{subfigure}
\begin{subfigure}[b]{0.25\textwidth}
    \centering
    \includegraphics[width=\linewidth, keepaspectratio]{images/br_evasion_attack_ex2_downloader.png}
    \caption{E3: Downloader}
    \label{fig:br_evasion_attack_ex3_downloader}
\end{subfigure}
\begin{subfigure}[b]{0.25\textwidth}
    \centering
    \includegraphics[width=\linewidth, keepaspectratio]{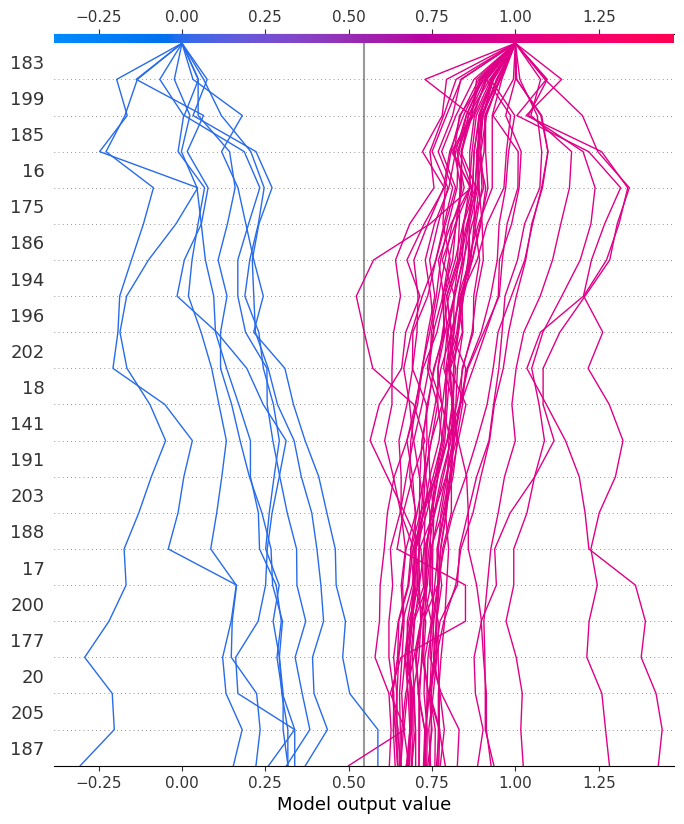}
    \caption{E1: Ransomware}
    \label{fig:br_evasion_attack_ex1_ransomware}
\end{subfigure}
\begin{subfigure}[b]{0.25\textwidth}
    \centering
    \includegraphics[width=\linewidth, keepaspectratio]{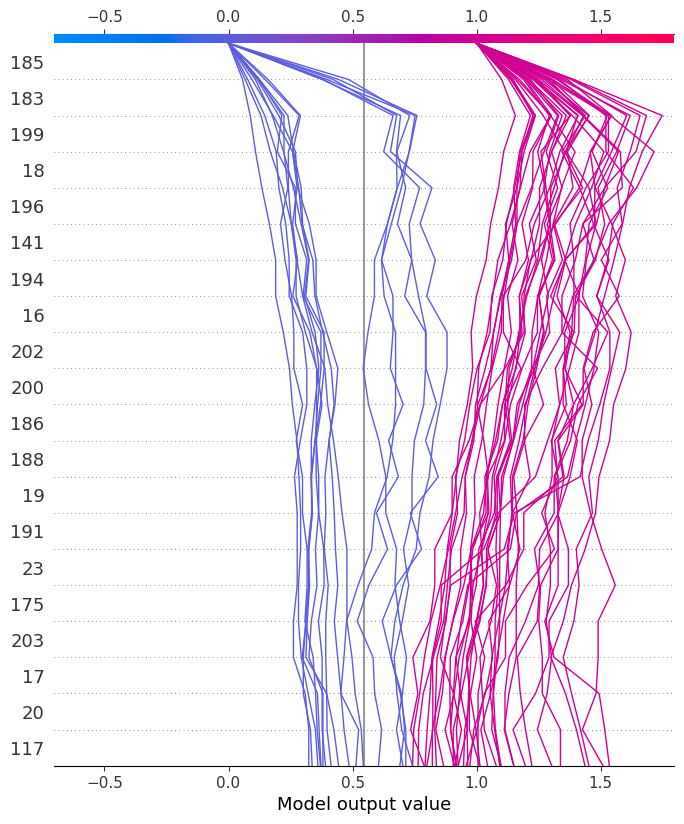}
    \caption{E2: Ransomware}
    \label{fig:br_evasion_attack_ex2_ransomware}
\end{subfigure}
\begin{subfigure}[b]{0.25\textwidth}
    \centering
    \includegraphics[width=\linewidth, keepaspectratio]{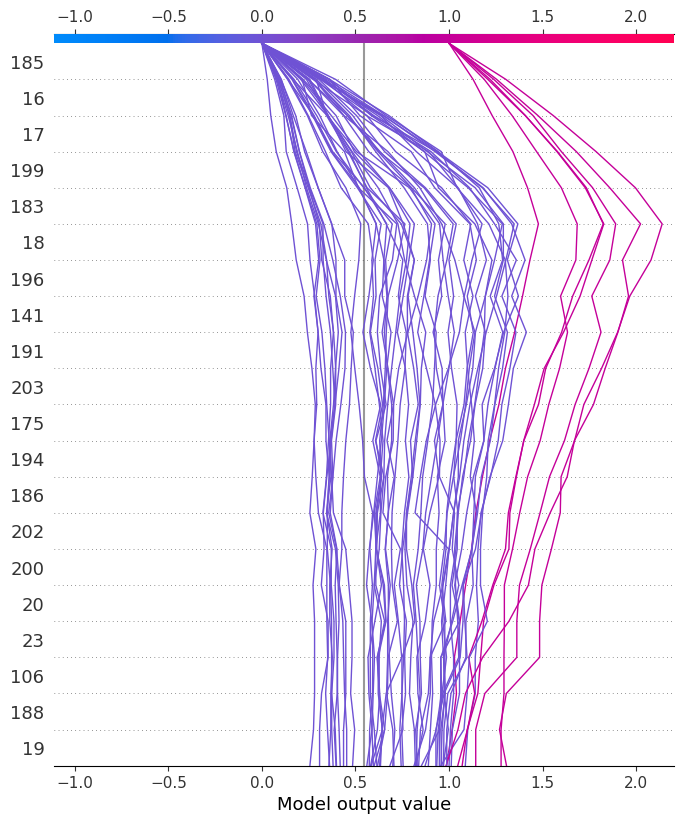}
    \caption{E3: Ransomware}
    \label{fig:br_evasion_attack_ex3_ransomware}
\end{subfigure}
\caption{Decision plots for evasion attack experiments on the BR technique}
\label{fig:br_evasion_attack_decision_plot}
\end{figure*}

Figure \ref{fig:br_evasion_attack_decision_plot} shows the decision plots for the evasion attack experiment conducted in Section \ref{sec:rq3}. Figures \ref{fig:br_evasion_attack_ex1_downloader} and \ref{fig:br_evasion_attack_ex1_ransomware} show the original predictions for the 54 Ransomware-only samples. In Figures \ref{fig:br_evasion_attack_ex2_downloader} and \ref{fig:br_evasion_attack_ex3_downloader}, we can see how adversarial perturbations can change any negative predictions to positive, while Figures \ref{fig:br_evasion_attack_ex2_ransomware} and \ref{fig:br_evasion_attack_ex3_ransomware} show the vice versa.


\end{document}